\documentclass[floatfix,11pt,nofootinbib]{revtex4}

\usepackage{amssymb,amsmath}
\usepackage{graphicx,float}
\usepackage{indentfirst}
\def \pt{\partial}

\def \O{\mathcal{O}}
\def \D{\mathcal{D}}

\def \Im{\textmd{Im}}
\def \and{\textmd{and}}
\def \nn{\nonumber}

\begin{document}

\title{\Large{\boldmath Interesting features of a general class of higher-derivative theories of quantum gravity}}

\author{A. Accioly} \email{accioly@cbpf.br} 
\author{J. de Almeida} \email{josejr@cbpf.br}
\author{G.P. de Brito} \email{gpbrito@cbpf.br}
\author{W. Herdy} \email{wallacew@cbpf.br}

\affiliation{Centro Brasileiro de Pesquisas F\'{i}sicas (CBPF),\\ Rua Dr Xavier Sigaud 150, Urca, Rio de Janeiro, RJ, Brazil, CEP 22290-180}

\begin{abstract}
In this work we investigate an interesting connection between the absence of Newtonian singularities in the classical non-relativistic potential and renormalizability properties in Higher-Derivative models of Quantum Gravity.  In the framework of a large class of $D$-dimensional Higher-Derivative models of Quantum Gravity, we compute the non-relativistic potential energy associated with two point-like masses. Investigating its behavior for small distances, we find an algebraic condition which is sufficient for the cancellation of the Newtonian singularity. We verify that the same condition is necessary to ensure power-counting renormalizability and, as a consequence,  we conclude that renormalizable Higher-Derivative models do not exhibit the so-called newtonian singularity. Finally, we discuss the role of ghosts in the mechanism for the cancellation of Newtonian singularities.	
\end{abstract}
\maketitle

\section{Introduction \label{Introduction}}

The quest for a quantum gravity theory is still  one of the most important problems of theoretical physics.  As is well  known, the biggest challenge in the construction of a quantum theory for the gravitational interaction is the lack of experimental evidences concerning  gravity at the microscopic level. Despite that, however, there exist several approaches to quantum gravity which were  proposed in the last few decades. For instance: string theory, loop quantum gravity, causal dynamical triangulations, causal sets and induced quantum gravity \cite{Kiefer,Oriti}. Nevertheless,  none of the aforementioned theories can be considered a complete quantum gravity theory up to now. 

At the classical level, the gravitational interaction is very well described in terms of Einstein's general relativity (GR), which is confirmed for  the excellent  concordance between its theoretical predictions  and the  available experimental  tests (e.g. solar system tests, cosmological observations and the recent discovery  of gravitational waves). Therefore, a natural path for the construction of a quantum theory of gravity seems, at least at first sight, the quantization of GR. 

Now, keeping in mind that the fundamental field concerning  GR is the spacetime metric, a possible way to perform its quantization entails a functional integration over all metric fluctuations around some vacuum configuration \cite{Kiefer,Shapiro}. For instance, we may consider the metric splitting $g_{\mu\nu} = \eta_{\mu\nu} + \kappa h_{\mu\nu}$ and  perform afterward the path integral quantization of the fluctuation field $h_{\mu\nu}$. In this approach we can derive at the tree level the Feynman rules of the theory and therefore apply the standard techniques of perturbative QFT. An interesting result that can be obtained via the semiclassical approach is the gravitational light bending \cite{Donoghue}. Nevertheless, computations at the loop level are problematic.  In this case, the UV divergences are not treatable by means of perturbative renormalization and, as a consequence, the theory is UV incomplete.

A possible way out of  the UV divergences problem is the introduction of higher-derivative terms in the usual Einstein-Hibert action\footnote{An alternative route to deal with that problem is the so-called Asymptotic Safety program \cite{Percacci}. On the other hand, within the context of Asymptotic Safe Quantum Gravity the problem of UV divergences may be carried out via the use of nonperturbative renormalization. In this paper, we shall restrict ourselves to perturbative techniques.}. In fact, these  higher-derivative terms improve the behavior of the tree-level propagator in such a way that it compensates the ``nasty'' UV behavior of the vertices containing derivative couplings. In a seminal paper \cite{Stelle:1976gc}, Stelle investigated thoroughly  the renormalizability of higher- derivative theories of quantum gravity and came to the conclusion  that  a fourth- derivative theory described by an action containing curvature-squared terms is renormalizable (in 4-dimensions) to all orders of perturbation theory. However, there is a price to be paid for attaining renormalizability: the spectra of the theory exhibits a massive ghost-like particle which can cause unitarity violation. In addition, from the classical point of view, higher derivatives may lead to Ostrogradsky's instabilities. Unfortunately, nonunitarity is a problem as undesirable as nonrenormalizability. Therefore,  we have to found a  way to circumvent it if we want to follow the route of higher-derivative theories of quantum gravity\footnote{Remarkably, an Euclidean lattice formulation of the fourth -derivative quantum gravity points out that unitarity may be restored at nonperturbative level \cite{Tomboulis}.}.

In the last few decades many efforts have been made to reconcile unitarity and renormalizability of higher-derivative theories. For instance, it was verified that if we allow terms with six or higher derivatives in the action, then the theory becomes superrenormalizable \cite{Shapiro:1997,Modesto:2012,Modesto:2012ys} and it is possible to find some region in the parameter space where all the nontrivial poles of the tree-level propagator are complex and, therefore, the theory may be formulated as unitary in the Lee-Wick sense \cite{Modesto:2016,Shapiro:2016}. We remark, however, that the presence of complex poles may lead to problems with causality \cite{Yamamoto:1969}. Another possibility arises if we allow nonlocal terms in the action \cite{Biswas:2012,Modesto:2014,Modesto:2017}. In this case it is possible to choose a form factor such that the only pole of the propagator corresponds to the usual physical graviton and, as a consequence, we may escape from the unitarity violation. Besides, a typical nonlocal term improves the tree-level propagator in such a way that the theory becomes super-renormalizable or even UV finite\footnote{It is important to emphasize that there are some classes of nonlocal theories where both renormalizability and tree-level unitarity are respected; however, after quantum corrections are introduced it is possible that they become nonunitary \cite{Shapiro:2015}.}. In the past few years the study of non-local theories of quantum gravity has been applied to several physical situations, \textit{e.g.} cosmological scenarios and astrophysical properties \cite{Biswas:2005,Koshelev:2017}, high energy scatterings \cite{Talaganis:2016} and so on. 

If we restrict ourselves to local gravitational theories, however, a pacific coexistence between renormalizability and unitarity is generally not attained. A conjecture first proposed by Stelle \cite{Stelle:1976gc} that renormalizable higher-order gravity models are endowed with a classical potential lacking a singularity at the origin allows us to look at this incompatibility through a different lens. Since a unitary system is correlated to a singular potential at the origin \cite{Accioly:2013hwa}, while a renormalizable model is related to a potential finite at the origin, this conjecture has as a consequence the impossibility of having higher order gravitational theories that are simultaneously unitary and renormalizable. 

Recently, many authors have addressed the problem of verifying this hypothesis for several models. It has been verified in $D$-dimensions for fourth and sixth-derivative derivative gravity \cite{Accioly:2017-1,Accioly:2017-2} and also for scale invariant gravity \cite{Scaleinvariant:2017}. The converse of this hypothesis was shown to be false, in other words, that a higher-order gravitational model which has a finite classical potential at the origin is not necessarily renormalizable \cite{Breno:2016}. 

In this paper we intend to probe this conjecture for a general class of $D$-dimensional higher-derivative gravity theories. In this vein, we compute a general expression for the interparticle potential energy and analyze its behavior for small distances. A sufficient condition for the regularity of the potential at small distances is then found and, as we shall see, this condition turns out to be automatically satisfied by power-counting renormalizable theories. For the sake of completeness, we shall discuss the tree-level unitarity of the this general class and verify that the theories which are ghost-free turns out to be power counting nonrenormalizable. In addition, the role of ghost-like particles on the cancellation of the newtonian singularity is discussed as well \cite{Tiberio:2015,Breno:2016}.

This paper is organized as follows: In section \ref{General_class}, we present a general class of higher-derivative gravity models and investigate some of its properties. In the section \ref{Potential} we compute a general expression for the interparticle potential and discuss its behavior for small distances. In section \ref{Renormalizability}, we study the UV behavior of the system and its renormalizability properties. In section \ref{unitarity_section}, we analyze the particle spectra of the theory as well as its implications for tree-level unitarity. In section \ref{Conjecture}, we summarize our findings and probe the conjecture that the absence of Newtonian singularities is a necessary condition for these theories to be renormalizable. Finally, in section \ref{conclusion} we present our conclusions.

Throughout this paper we use the conventions $c = \hbar = 1$, $\eta_{\mu\nu} = diag(+,-,\cdots,-)$, $R^{\mu}_{\,\,\,\nu\alpha\beta} = \pt_{\alpha}\Gamma^{\mu}_{\nu\beta} + \Gamma^{\mu}_{\alpha\lambda}\Gamma^\lambda_{\nu\beta} - (\alpha \leftrightarrow \beta)$, $R_{\mu\nu} = R^\beta_{\,\,\,\mu\nu\beta}$ and $R=g^{\mu\nu}R_{\mu\nu}$.

\section{General class of higher-derivative gravity models \label{General_class}}

Let us start considering  a general class of higher-derivative gravity models.  Although the gravitational action has an infinite number of terms that are compatible  with the symmetries under general coordinate transformations, we shall only  consider the quadratic sector of the action\footnote{Naturally the discussion of renormalizability takes into account non-quadratic terms. However the specific form of these contributions will not be relevant for our purposes.}. In this spirit, the most general $D$-dimensional ($D\geq 3$) action is given by
\begin{eqnarray}\label{action_general}
S[g_{\mu\nu}] = \int d^Dx \sqrt{|g|}\bigg( \frac{2\sigma}{\kappa^2} R + \frac{1}{2\kappa^2}R F_1(\Box) R + \frac{1}{2\kappa^2} R_{\mu\nu} F_2(\Box) R^{\mu\nu} \bigg),
\end{eqnarray}
where $\kappa^2 = 32\pi G = 32\pi M_{Pl}^{2-D}$ is the gravitational coupling constant, $\sigma$ is a parameter that can be taken to be either $+1$ or $-1$ and $F_1(\Box)$ and $F_2(\Box)$ are functions of the covariant d'Alembertian operator ($\Box = g^{\mu\nu}\nabla_\mu \nabla_\nu$). Moreover, we shall assume that the functions $F_1(\Box)$ and $F_2(\Box)$, the so-called \textit{form factors}, have a finite polynomial representation.

At this point some comments regarding the action above are in order:
\begin{itemize}
	\item As it was mentioned before, our goal is to consider the most general contribution for the quadratic sector of the gravitational action. In this
    sense, there are several non-trivial steps before equation \eqref{action_general}. As it was argued in references \cite{Biswas:2012,Biswas:2013}, the most general 
    contribution involving quadratic curvature terms takes the following form
    \begin{eqnarray}
    \int d^Dx \sqrt{|g|} \, R_{\mu\nu\alpha\beta} \mathcal{D}^{\mu\nu\alpha\beta}_{\rho\lambda\sigma\gamma} R^{\rho\lambda\sigma\gamma} ,
    \end{eqnarray}
    where $\mathcal{D}^{\mu\nu\alpha\beta}_{\rho\lambda\sigma\gamma}$ is a differential operator constructed with all possible combinations involving covariant
    derivatives and the spacetime metric. At first sight, this contribution can be written in terms of fourteen invariant terms (see Eq. (3) in reference
    \cite{Biswas:2013}). After some rearrangements based in the use of the antisymmetric properties of the Riemann tensor and also using the Jacobi identity, the most
    general curvature quadratic contribution can be recast in terms of six independent operators, namely
    \begin{align}
    &\int \!d^Dx \sqrt{|g|} \,\Big( R F_1(\Box)R \!+\! R_{\mu\nu} F_2(\Box)R^{\mu\nu} \!+\! R_{\mu\nu\alpha\beta} F_3(\Box) R^{\mu\nu\alpha\beta} \!+\! R
    F_4(\Box) \nabla_\mu \nabla_\nu \nabla_\alpha \nabla_\beta R^{\mu\nu\alpha\beta}  + \nn \\ 
    &\!+
    R_{\mu}^{\,\,\,\,\lambda\rho\sigma} F_5(\Box) \nabla_\lambda \nabla_\rho \nabla_\sigma \nabla_\nu \nabla_\alpha \nabla_\beta R^{\mu\nu\alpha\beta} 
    \!+\! R^{\lambda\rho\sigma\gamma} F_6(\Box) \nabla_\lambda \nabla_\rho \nabla_\sigma \nabla_\gamma \nabla_\mu \nabla_\nu \nabla_\alpha \nabla_\beta R^{\mu
    \nu\alpha\beta}\Big).
    \end{align}
    Since we are interested in considering only quadratic contributions with respect to $h_{\mu\nu}$, we can replace those covariant derivatives appearing in the
    last expression by ordinary derivative operators. In these conditions, the derivative operators commute and, as consequence, those terms associated with the form
    factors $F_4(\Box)$, $F_5(\Box)$ and $F_6(\Box)$ vanishes. It should be emphasized that some points of the above discussion relies on the assumption that we
    are dealing with fluctuations defined with respect to a Minkowskian background. For a discussion about higher-derivative models with (anti-)de Sitter
    background, we recommend \cite{Biswas_dS_AdS,Biswas_dS_AdS_2}. Also, it is important to mention that the classical dynamics associated with this general class of quadratic curvature gravity was studied within the full nonlinear regime in reference \cite{Biswas_Eq_Motion}. 
	\item Now we turn our attention to the contribution coming from the invariant term 
	\begin{eqnarray}
	R_{\mu\nu\alpha\beta} F_3(\Box)R^{\mu\nu\alpha\beta} .
	\end{eqnarray} 
	Indeed, this is a legitimate term from the point of view of spacetime symmetries and it apparently contributes to the quadratic sector. However, looking closer this is not completely true. In fact, taking into account small fluctuations around the Minkowskian background, $g_{\mu\nu} = \eta_{\mu\nu} + \kappa h_{\mu\nu}$, we arrive at the following result 
	\begin{eqnarray}\label{Gauss_Bonnet}
	R_{\mu\nu\alpha\beta} F_3(\Box)R^{\mu\nu\alpha\beta} = 4 R_{\mu\nu} F_3(\Box) R^{\mu\nu} - RF_3(\Box)R + \pt \Omega + \O(h^3) .
	\end{eqnarray} 
	Since we are mainly interested in the quadratic part of the action we can discard the contribution of $R_{\mu\nu\alpha\beta} F_3(\Box)R^{\mu\nu\alpha\beta}$ by a simple redefinition of the functions $F_1(\Box)$ and $F_2(\Box)$. To be precise, if we start with
	\begin{align}
	S[g_{\mu\nu}] =& \int d^Dx \sqrt{|g|}\bigg( \frac{2\sigma}{\kappa^2} R + \frac{1}{2\kappa^2}R F_1(\Box) R + \nn\\ &+ \frac{1}{2\kappa^2} R_{\mu\nu} F_2(\Box) R^{\mu\nu} + \frac{1}{2\kappa^2} R_{\mu\nu\alpha\beta} F_3(\Box)R^{\mu\nu\alpha\beta} \bigg)
	\end{align}
	and then consider equation \eqref{Gauss_Bonnet}, we may rewrite the last expression as follows
	\begin{align}
	S[g_{\mu\nu}] =& \int d^Dx \sqrt{|g|}\bigg[ \frac{2\sigma}{\kappa^2} R + \frac{1}{2\kappa^2}R \Big( F_1(\Box) - F_3(\Box) \Big) R + \nn\\ &+ \frac{1}{2\kappa^2}  R_{\mu\nu} \Big(F_2(\Box) + 4 F_3(\Box) \Big) R^{\mu\nu} \bigg] + \int d^Dx \,\pt \Omega ,
	\end{align}
	where we have removed the contribution $\O(h^3)$. Keeping  in mind  that $\int d^Dx \,\pt \Omega = 0$ with the proper boundary condition and using the following redefinitions
	\begin{eqnarray}
	F_1(\Box) - F_3(\Box) \mapsto F_1(\Box) \qquad \textmd{and}\qquad F_2(\Box) + 4 F_3(\Box) \mapsto F_2(\Box) 
	\end{eqnarray}
	we recover our original action \eqref{action_general}.
	\item As it was mentioned before, there are an infinity number of compatible terms with the symmetries of the gravitational interaction. These terms may be obtained through all possible invariant combinations of the Riemann tensor, the Ricci tensor and the Ricci scalar, \textit{e.g.} $R^3$, $R_{\mu\nu}R^{\mu\nu} R$, $R_{\mu\nu\alpha\beta}R^{\mu\alpha} R^{\nu\beta}$, $R^4$, $(R_{\mu\nu} R^{\mu\nu})^2$ and so on. However, it is not difficult to see that the aforementioned contributions bring only terms of order $\O(h^3)$, which are not relevant to the quadratic part of the action.
	\item Finally, we have introduced the constant parameter $\sigma$ in order to explore some interesting features related to the unitarity of 3-dimensional higher-derivative models. In addition, without loss of generality we will take this parameter to be either $\sigma = +1$ or $\sigma = -1$.
\end{itemize}

\section{Non-relativistic potential energy \label{Potential}}

In order to analyze the conjecture that relates renormalizability to the cancellation of Newtonian singularities, let us compute the interparticle potential energy for the general higher -derivative gravity theory given by ({\ref{action_general}}) and investigate its behavior for small distances. For this purpose we shall employ the prescription presented by Accioly \textit{et. al.}, which is based on the path integral formulation of quantum field theory \cite{Accioly-Prescrip}. This prescription states that, in order to determine the potential energy of gravitational models, we need only to compute
\begin{eqnarray}\label{prescription}
E_{D}(r)=\frac{\kappa^2}{4}\frac{M_1M_2}{(2\pi)^{D-1}}\int d^{D-1} \textbf{k} \, e^{ \textbf{k} \cdot \textbf{r}}\, P_{00,00}(k)|_{k_0=0},
\end{eqnarray}
where $P_{00,00}$ is the  $\mu = \nu = \alpha = \beta = 0$ component of the modified propagator $P_{\mu\nu,\alpha\beta}=D_{\mu\nu,\alpha\beta}-D_{\mu\nu,\alpha\beta}^{\perp}$, with $D_{\mu\nu,\alpha\beta}^{\perp}$ being the contribution to the propagator that is orthogonal to the energy-momentum tensor, while, as before, $\kappa^2=32\pi G.$

After a standard procedure \cite{Accioly:2000}, the free propagator (in momentum space) associated with the general higher-derivative gravity \eqref{action_general} may be cast as
\begin{align} \label{Propagator}
&\D_{\mu\nu,\alpha\beta}(k) = \frac{1}{\sigma k^2 Q_2(k^2)}P^{(2)}_{\mu\nu,\alpha\beta} - \frac{1}{(D-2)} \frac{1}{\sigma k^2 Q_0(k^2)}  P^{(0-s)}_{\mu\nu,\alpha\beta} + \frac{2 \lambda}{k^2} P^{(1)}_{\mu\nu,\alpha\beta} + \nn \\
&+  \bigg( \frac{4\lambda}{k^2} -  \frac{(D-1)}{\sigma(D-2) k^2 Q_0(k^2)} \bigg) P^{(0-w)}_{\mu\nu,\alpha\beta} - \frac{\sqrt{D-1}}{\sigma(D-2) k^2 Q_0(k^2)} \Big( P^{(0-sw)}_{\mu\nu,\alpha\beta} + P^{(0-ws)}_{\mu\nu,\alpha\beta} \Big) ,
\end{align}
where $\{ P^{(2)},\cdots,P^{(0-ws)} \}$ denotes the set of Barnes-Rivers operators, $\lambda$ is the gauge fixing parameter\footnote{We have considered the de Donder gauge condition in the computation  of the free propagator.} and we have defined
\begin{eqnarray}\label{Q_2}
Q_2(k^2) = 1 + \frac{1}{4\sigma} k^2 F_2(-k^2),
\end{eqnarray}
and
\begin{eqnarray}\label{Q_0}
Q_0(k^2) = 1 - \frac{k^2}{\sigma(D-2)} \bigg((D-1)F_1(-k^2) + \frac{D}{4} F_2(-k^2) \bigg).
\end{eqnarray}
It is important to emphasize that although we are dealing with polynomial form factors (\textit{i.e.} local theories), the propagator above reported is also valid for the case of non-local form factors and it can be compared with those results presented in \cite{Conroy,Biswas:2013}.

Now, since we have already determined the propagator in (\ref{Propagator}), it is straightforward to see that
\begin{eqnarray}
P_{00,00}(k)=\frac{D-2}{D-1}\frac{1}{\sigma k^2Q_2(k^2)}-\frac{1}{(D-1)(D-2)}\frac{1}{\sigma k^2Q_0(k^2)}.
\end{eqnarray}
As it was previously mentioned $F_1(\Box)$ and $F_2(\Box)$ are polynomial functions of the d'Alembertian operator, namely
\begin{eqnarray}\label{Polynomial_function}
F_1(\Box) = \sum_{n=0}^p \alpha_n (-\Box)^n   
\qquad \textmd{and} \qquad
F_2(\Box) = \sum_{n=0}^q \beta_n (-\Box)^n  , 
\end{eqnarray}
where $\alpha_n$ and $\beta_n$ are real coefficients with canonical mass dimension $M^{-2(n+1)}$. Let
\begin{eqnarray}\label{definitionm}
\{m_{(2),1}^2,m_{(2),2}^2,\cdots,m_{(2),\tilde{q}+1}^2\} \qquad \textmd{and}  \qquad \{m_{(0),1}^2,m_{(0),2}^2,\cdots,m_{(0),\tilde{N}+1}^2\},
\end{eqnarray}
be, respectively, the set of real roots of the polynomial functions $Q_2(k^2)$ and $Q_0(k^2)$, while
\begin{eqnarray}\label{definitioneta}
\{\eta_{(2),1}^2,\eta_{(2),1}^{*\,2},\cdots,\eta_{(2),r}^2,\eta_{(2),r}^{*\,2}\} \qquad \textmd{and}  \qquad \{\eta_{(0),1}^2,\eta_{(0),1}^{*\,2},\cdots,\eta_{(0),s}^2,\eta_{(0),s}^{*\,2}\},
\end{eqnarray}
denote, respectively, the sets of complex roots of the polynomials $Q_2(k^2)$ and $Q_0(k^2)$. Therefore, 
we arrive at the constraints
\begin{eqnarray}
q = \tilde{q} + 2 r \qquad \textmd{and} \qquad \textmd{max}\{p,q\} = \tilde{N} + 2s \equiv N.
\end{eqnarray}
By the factorization theorem for polynomials and partial fraction decomposition one may write \eqref{Propagator} as follows
\begin{align}\label{modifiedpropagatoralgebra}
P_{00,00}(k)|_{k_0=0}&=-\frac{1}{\sigma}\left(\frac{D-3}{D-2}\right)\frac{1}{\textbf{k}^2}-\frac{1}{\sigma}\left(\frac{D-2}{D-1}\right) \sum_{i=1}^{q+1} \prod_{\substack{j = 1\\ j \neq i}}^{q+1} \!\frac{ \mu_{(2),j}^2 }{\mu_{(2),j}^2 \!-\! \mu_{(2),i}^2} \frac{1}{\textbf{\textbf{k}}^2+\mu^2_{(2),i}}+ \nn \\
&+\frac{1}{\sigma}\frac{1}{(D-1)(D-2)}\sum_{i=1}^{N+1} \!\prod_{\substack{j = 1\\ j \neq i}}^{N+1} \!\frac{ \mu_{(0),j}^2 }{\mu_{(0),j}^2 \!-\! \mu_{(0),i}^2} \frac{1}{\textbf{k}^2+\mu^2_{(0),i}},
\end{align}
where we have defined
\begin{align}
\mu_{(2),i} = \begin{cases}
m_{(2),i} \quad \, , i = 1,\cdots, \tilde{q}+1 ,\\
\eta_{(2),i} \quad \, , i = \tilde{q}+ 2,  \cdots, \tilde{q} + r +1 , \\
\eta^*_{(2),i} \quad \, , i = \tilde{q}+ r + 2,  \cdots, \tilde{q} + 2 r +1 , 
\end{cases}
\end{align}
and
\begin{align}
\mu_{(0),i} = \begin{cases}
m_{(0),i} \quad \, , i = 1,\cdots, \tilde{N}+1 ,\\
\eta_{(0),i} \quad \, , i = \tilde{N}+ 2,  \cdots, \tilde{N} + s +1 , \\
\eta^*_{(0),i} \quad \, , i = \tilde{N}+ s + 2,  \cdots, \tilde{N} + 2 s +1 .
\end{cases} 
\end{align}
Substituting (\ref{modifiedpropagatoralgebra}) into (\ref{prescription}) and taking into account the integrals
\begin{subequations}
	\begin{eqnarray}
	\int \frac{d^{D-1}\textbf{k}}{(2\pi)^{D-1}}\frac{e^{i\textbf{k}\cdot\textbf{r}}}{\textbf{k}^2+\mu^2}=\frac{1}{(2\pi)^{\frac{D-1}{2}}}\left(\frac{\mu}{r}\right)^{\frac{D-3}{2}}K_{\frac{D-3}{2}}(\mu r), \qquad \text{ for } D \geq \text{3}
	\end{eqnarray}
	
	\begin{eqnarray}
	\int \frac{d^{D-1}\textbf{k}}{(2\pi)^{D-1}}\frac{e^{i\textbf{k}\cdot\textbf{r}}}{\textbf{k}^2}=\frac{1}{(2\pi)^{\frac{D-1}{2}}}\frac{2^{\frac{D-5}{2}}}{r^{D-3}}\Gamma \left( \frac{D-3}{2}\right),\qquad \text{ for } D \geq \text{4},
	\end{eqnarray}
\end{subequations}
we find that the $D$-dimensional gravitational potential energy is given by (for $D \geq 4$)
\begin{align}\label{potencial}
E_D(r)=&-\frac{\kappa^2 M_1M_2}{4\sigma(2\pi)^\frac{D-1}{2}}\Bigg\{\left( \frac{D-3}{D-2}\right)2^{\frac{D-5}{2}}\Gamma\left( \frac{D-3}{2}\right) \frac{1}{r^{D-3}}+\nn \\
&- \left(\frac{D-2}{D-1}\right) \sum_{i=1}^{q+1} \prod_{\substack{j = 1\\ j \neq i}}^{q+1} \!\frac{ \mu_{(2),j}^2 }{\mu_{(2),j}^2 \!-\! \mu_{(2),i}^2} \bigg(\frac{\mu_{(2),i}}{r}\bigg)^{\frac{D-3}{2}}K_{\frac{D-3}{2}}(\mu_{(2),i}r) + \nn\\ 
&+\frac{1}{(D-1)(D-2)} \sum_{i=1}^{N+1} \!\prod_{\substack{j = 1\\ j \neq i}}^{N+1} \!\frac{ \mu_{(0),j}^2 }{\mu_{(0),j}^2 \!-\! \mu_{(0),i}^2} \bigg(\frac{\mu_{(0),i}}{r}\bigg)^{\frac{D-3}{2}}K_{\frac{D-3}{2}}(\mu_{(0),i}r)  \Bigg\}.
\end{align}
Similarly, for $D=3$ the interparticle gravitational potential energy is determined to be 
\begin{align}\label{potencial_d3}
E_3(r)=&\frac{\kappa^2 M_1M_2}{8\sigma(2\pi)}\Bigg\{ \sum_{i=1}^{q+1} \prod_{\substack{j = 1\\ j \neq i}}^{q+1} \!\frac{ \mu_{(2),j}^2 }{\mu_{(2),j}^2 \!-\! \mu_{(2),i}^2} K_{0}(m_{(2),i}r) +\nn \\&- \sum_{i=1}^{N+1}\prod_{\substack{j = 1\\ j \neq i}}^{N+1} \!\frac{ \mu_{(0),j}^2 }{\mu_{(0),j}^2 \!-\! \mu_{(0),i}^2} K_{0}(m_{(0),i}r)  \Bigg\} .
\end{align}

We remark that the expression for the non-relativistic potential energy of the same class of higher-derivative models that we are considering here was already computed in the particular case of $D=4$ \cite{Tiberio:2015,Breno:2016}. Also, we note that reference \cite{Quandt} presents the non-relativistic potential energy obtained for higher-order gravities containing the Ricci scalar sector. It is important to emphasize that our results agree with both of them when proper limits are taken.

Our next step will be to analyze the behavior of this gravitational potential energy for small distances. Defining $\nu=\frac{D-3}{2}$, we shall make a distinction between $D$ odd and even, since a modified Bessel function of the second kind $K_{\nu}(x)$ has a Taylor's series expansion which depends if $\nu$ is integer or half-integer. 

\subsection{Regularity of the potential energy at the origin for $D > 3$ - $D$ even}

If the potential energy is defined in a spacetime with even dimensions, we can expand the modified Bessel function of the second kind according to
\begin{eqnarray}\label{expansaoBessellpar}
K_\nu(z) = \frac{\pi\csc(\pi\nu)}{2}\left(\sum_{k=0}^{\infty}\frac{1}{\Gamma(k-\nu+1)k!}\left(\frac{z}{2}\right)^{2k-\nu}-\sum_{k=0}^{\infty}\frac{1}{\Gamma(k+\nu+1)k!}\left(\frac{z}{2}\right)^{2k+\nu}\right),\nonumber
\end{eqnarray}
and, substituting the above expression in \eqref{potencial}, we  determine the gravitational potential energy for small distances to be
\begin{align}\label{potencial_expansao_par}
&E_D(r)=-\frac{\kappa^2 M_1M_2}{4\sigma(2\pi)^{\nu+1}}\Bigg\{ \frac{1}{r^{2\nu}}\Delta^{\textmd{even}}_\nu(r;q,N) +\frac{\pi\csc(\pi\nu)}{2^{\nu+1}\Gamma (\nu+1)}\Bigg[ \left( \frac{2\nu+1}{2\nu+2}\right)\times \nn \\
&\times \sum_{i=1}^{q+1} \prod_{\substack{j = 1\\ j \neq i}}^{q+1} \frac{ \mu_{(2),j}^2 }{\mu_{(2),j}^2 - \mu_{(2),i}^2} \, \mu_{(2),i}^{2 \nu}  -  \frac{1}{(2\nu+2)(2\nu+1)}\sum_{i=1}^{N+1} \prod_{\substack{j = 1\\ j \neq i}}^{N+1} \frac{ \mu_{(0),j}^2 }{\mu_{(0),j}^2 - \mu_{(0),i}^2} \, \mu_{(0),i}^{2 \nu}\Bigg]\Bigg\} +\mathcal{O}(r),
\end{align}
where
\begin{align}\label{delta_par}
&\Delta^{\textmd{even}}_\nu(r;q,N) =\frac{2^{\nu} \,\Gamma(\nu +  1)}{2\nu + 1}  - \pi\csc{(\pi\nu)}\sum_{k=0}^{\nu-\frac{1}{2}}\frac{r^{2k}}{2^{2k-\nu+1}\Gamma (k-\nu+1)k!}\times\nn\\
&\times\left( \frac{2\nu+1}{2\nu+2} \sum_{i=1}^{q+1} \prod_{\substack{j = 1\\ j \neq i}}^{q+1} \frac{ \mu_{(2),j}^2 }{\mu_{(2),j}^2 - \mu_{(2),i}^2} \, \mu_{(2),i}^{2 k} - \frac{1}{(2\nu+2)(2\nu+1)}\sum_{i=1}^{N+1} \prod_{\substack{j = 1\\ j \neq i}}^{N+1} \frac{ \mu_{(0),j}^2 }{\mu_{(0),j}^2 - \mu_{(0),i}^2} \, \mu_{(0),i}^{2 k}\right)\nn.
\end{align}

Therefore, the cancellation of  Newtonian singularity depends on the behavior of the function $\Delta^{\textmd{even}}_\nu(r;q,N)$. With the help of a computer algebra system we may verify that this function will be null for all values of $r$ if the following condition is satisfied,
\begin{eqnarray}
2q + 4 -D \geq 0.
\end{eqnarray}

At first glance this result might seem surprising, after all, we are concluding that the finiteness of the potential near the origin is independent of the scalar curvature squared sector. We shall discuss in the section \ref{Conjecture} the reasons for this being so. It is also worthwhile to note that we could obtain finite results at $r=0$ even if the above condition is not met. However, we would need to adjust the parameters of the Lagrangian.

\subsection{Regularity of the potential energy at the origin for $D >3$ - $D$ odd}

For a spacetime with odd dimensions and $D > 4$, we can expand the modified Bessel function of the second kind according to
\begin{align}
K_\nu(z) &= (-1)^{\nu -1} \ln\left( \frac{z}{2} \right) \sum_{k=0}^{\infty} \frac{1}{k!(k+\nu)!} \left( \frac{z}{2} \right)^{\nu + 2k} + \frac{1}{2} \left(\frac{2}{z}\right)^\nu \sum_{k=0}^{\nu-1} \frac{(-1)^k (\nu -k -1)! }{k!} \left( \frac{z}{2} \right)^{2k} +\nn \\
&+ \frac{(-1)^\nu}{2} \sum_{k=0}^{\infty} \frac{\psi(k+1) + \psi(k+\nu+1)}{k!(k+\nu)!} \left( \frac{z}{2} \right)^{\nu+ 2k},
\end{align}
and, substituting the above expression in \eqref{potencial}, we found that the gravitational potential energy for small distances is 
\begin{align}\label{potencial_expansao_impar}
&E_D(r)=-\frac{\kappa^2 M_1M_2}{4\sigma(2\pi)^\frac{D-1}{2}}\Bigg\{  \frac{1}{r^{2\nu}}\Delta_\nu^{\textmd{odd}}(r;q,N) + \frac{(-1)^{\nu+1}}{2^{\nu+1} \, \nu!} \Bigg[ \Big( \psi(1) + \psi(\nu+1) \Big) \times \nn\\
&\times \bigg(\frac{2\nu +1}{2\nu+2}  \sum_{i=1}^{q+1} \prod_{\substack{j = 1\\ j \neq i}}^{q+1} \!\frac{ \mu_{(2),j}^2 }{\mu_{(2),j}^2 \!-\! \mu_{(2),i}^2} \, \mu_{(2),i}^{2 \nu}  - \frac{1}{(2\nu+2)(2\nu+1)}  \sum_{i=1}^{N+1} \prod_{\substack{j = 1\\ j \neq i}}^{N+1} \!\frac{ \mu_{(0),j}^2 }{\mu_{(0),j}^2 \!-\! \mu_{(0),i}^2} \, \mu_{(0),i}^{2 \nu} \bigg) + \nn \\
&-\frac{2\nu +1}{2\nu+2}  \sum_{i=1}^{q+1} \prod_{\substack{j = 1\\ j \neq i}}^{q+1} \!\frac{ \mu_{(2),j}^2 }{\mu_{(2),j}^2 \!-\! \mu_{(2),i}^2} \, \mu_{(2),i}^{2 \nu} \,\ln(\mu_{(2),i}^2)  +\nn \\
&+ \frac{1}{(2\nu+2)(2\nu+1)} \bigg( \sum_{i=1}^{N+1} \prod_{\substack{j = 1\\ j \neq i}}^{N+1} \!\frac{ \mu_{(0),j}^2 }{\mu_{(0),j}^2 \!-\! \mu_{(0),i}^2} \, \mu_{(0),i}^{2 \nu}\,\ln(\mu_{(0),i}^2)  \Bigg] + \mathcal{O}(r) \Bigg\}, 
\end{align}
where we have defined
\begin{align}
&\Delta^{\textmd{odd}}_\nu(r;q,N) =\frac{2^{\nu} \,\Gamma(\nu +  1)}{2\nu + 1}  - \sum_{k=0}^{\nu-1} \frac{(-1)^k \, 2^{\nu-1}\,(\nu -k -1)! }{ k!} \left(\frac{r}{2} \right)^{2k} \times \nn\\
&\times \bigg(\frac{2\nu +1}{2\nu+2}  \sum_{i=1}^{q+1} \prod_{\substack{j = 1\\ j \neq i}}^{q+1} \!\frac{ \mu_{(2),j}^2 }{\mu_{(2),j}^2 \!-\! \mu_{(2),i}^2} \, \mu_{(2),i}^{2 k}  - \frac{1}{(2\nu+2)(2\nu+1)}  \sum_{i=1}^{N+1} \prod_{\substack{j = 1\\ j \neq i}}^{N+1} \!\frac{ \mu_{(0),j}^2 }{\mu_{(0),j}^2 \!-\! \mu_{(0),i}^2} \, \mu_{(0),i}^{2 k} \bigg) \nn \\ 
&+ \frac{(-1)^{\nu}}{2^{\nu} \, \nu!} r^{2\nu} \ln \left( \frac{r}{2} \right)\, \bigg(\frac{2\nu +1}{2\nu+2}  \sum_{i=1}^{q+1} \prod_{\substack{j = 1\\ j \neq i}}^{q+1} \!\frac{ \mu_{(2),j}^2 }{\mu_{(2),j}^2 \!-\! \mu_{(2),i}^2} \, \mu_{(2),i}^{2 \nu}  +\nn\\
&- \frac{1}{(2\nu+2)(2\nu+1)}  \sum_{i=1}^{N+1} \prod_{\substack{j = 1\\ j \neq i}}^{N+1} \!\frac{ \mu_{(0),j}^2 }{\mu_{(0),j}^2 \!-\! \mu_{(0),i}^2} \, \mu_{(0),i}^{2 \nu} \bigg). 
\end{align}

Analogously to the above section, the finiteness of  potential energy at $r=0$ will depend on the condition $\Delta^{\textmd{odd}}_\nu(r;q,N) = 0$ for every value of the coordinate $r$. With the help of a computer algebra system we can verify that a sufficient condition to get $\Delta^{\textmd{odd}}_\nu(r;q,N) = 0$ is given by
\begin{eqnarray}
2q + 3 - D \geq 0,
\end{eqnarray}
and the same conclusion obtained in the above section is valid.

\subsection{Regularity of the potential energy at the origin for $D = 3$}

Now let us investigate the $3$-dimensional case. Considering the expansion
\begin{eqnarray}
K_0(z) = -\ln \left( \frac{z}{2} \right) - \gamma + \mathcal{O}(z^2),
\end{eqnarray}
where $\gamma$ is the Euler-Mascheroni constant, it is straightforward to see that, for $D=3$, the gravitational potential energy for small distances is given by
\begin{align}\label{potential_D3}
E_3(r) =&- \frac{\kappa^2 M_1M_2}{16\pi \sigma }\Bigg\{ \Bigg[ \sum_{i=1}^{q+1} \prod_{\substack{j = 1\\ j \neq i}}^{q+1} \!\frac{ \mu_{(2),j}^2 }{\mu_{(2),j}^2 \!-\! \mu_{(2),i}^2}  - \sum^{N+1}_{i=1} \prod_{\substack{j = 1\\ j \neq i}}^{N+1} \!\frac{ \mu_{(0),j}^2 }{\mu_{(0),j}^2 \!-\! \mu_{(0),i}^2} \Bigg] \bigg( \ln \left( \frac{r}{2} \right) + \gamma  \bigg)  +\nn\\
&+\sum_{i=1}^{q+1} \prod_{\substack{j = 1\\ j \neq i}}^{q+1} \!\frac{ \mu_{(2),j}^2 }{\mu_{(2),j}^2 \!-\! \mu_{(2),i}^2} \ln \mu_{(2),i} - \sum^{N+1}_{i=1} \prod_{\substack{j = 1\\ j \neq i}}^{N+1} \!\frac{ \mu_{(0),j}^2 }{\mu_{(0),j}^2 \!-\! \mu_{(0),i}^2} \ln \mu_{(0),i}  + \mathcal{O}(r)  \Bigg\} .
\end{align}
In such a case we can use the identity\footnote{For a rigorous proof of this identity we refer to \cite{Breno:2016}.}
\begin{eqnarray}
\sum_{i=1}^{n+1} \prod_{\substack{j = 1\\ j \neq i}}^{n+1} \!\frac{ a_j }{a_j \!-\! a_i} = 1,
\end{eqnarray}
valid for any set of complex numbers $\{a_1,a_2,\cdots,a_{n+1}\}$ with $n\geq 0$, in order to verify the following equations
\begin{eqnarray}
\sum_{i=1}^{q+1} \prod_{\substack{j = 1\\ j \neq i}}^{q+1} \!\frac{ \mu_{(2),j}^2 }{\mu_{(2),j}^2 \!-\! \mu_{(2),i}^2} =1 \qquad \textmd{and} \qquad \sum^{N+1}_{i=1} \prod_{\substack{j = 1\\ j \neq i}}^{N+1} \!\frac{ \mu_{(0),j}^2 }{\mu_{(0),j}^2 \!-\! \mu_{(0),i}^2} = 1,
\end{eqnarray}
for $q \geq 0$ (note that $q\geq 0$ automatically implies $N\geq0$). Using the result above we can recast the interparticle potential energy for small distances as follows,
\begin{align}
E_3(r) =- \frac{\kappa^2 M_1M_2}{16\pi \sigma }\Bigg\{ \sum_{i=1}^{q+1} \prod_{\substack{j = 1\\ j \neq i}}^{q+1} \!\frac{ \mu_{(2),j}^2  \ln \mu_{(2),i}}{\mu_{(2),j}^2 -\mu_{(2),i}^2} -  \sum^{N+1}_{i=1} \prod_{\substack{j = 1\\ j \neq i}}^{N+1} \!\frac{ \mu_{(0),j}^2  \ln \mu_{(0),i}}{\mu_{(0),j}^2 - \mu_{(0),i}^2}  + \mathcal{O}(r)  \Bigg\}.
\end{align}
Therefore, a sufficient condition for cancellation of the Newtonian singularity in $3$-dimensional higher -derivative gravity is given by $q\geq 0$. In other words, this condition tells us that the existence of  Ricci squared sector is sufficient for the cancellation of Newtonian singularities.

Finally, it is important to emphasize that the above discussion is valid only for higher -derivative theories with unrelated parameters, otherwise the conditions for the cancellation of Newtonian singularities are not  sufficient. The so-called New Massive Gravity \cite{NMG} is a good example of this statement. In this theory the parameters are constrainted by $3\alpha_0 + 8\beta_0 = 0 $ (and $\sigma = -1$) and the interparticle potential energy for the NMG may be written as 
\begin{eqnarray}
E_{\textmd{NMG}}(r) = - \frac{\kappa^2 M_1 M_2}{16\pi} K_0(m_{(2),1} r) ,
\end{eqnarray}   
where $m_{(2),1}^2 = 4/\beta_0$. Taking into account the expansion of  Bessel function for small arguments we arrive at the  result
\begin{eqnarray}
E_{\textmd{NMG}}(r) = \frac{\kappa^2 M_1 M_2}{16\pi} \bigg[ \gamma + \ln\bigg( \frac{m_{(2),1} r}{2} \bigg) \bigg] \!+\! \O(r).
\end{eqnarray}
Therefore, we  have found a divergent potential energy at $r=0$. In the section \ref{Conjecture} we will discuss the mechanism for cancellation of Newtonian singularities which will clearly show  why this  phenomenon  does not occur in NMG theory. 

\subsection{Plotting results}

In order to complete the discussion of the previous sections let us analyze the graphical behavior of the non-relativistic potential energy calculated above. Since we want to deal with plots associated with dimensionless quantities, let us define a reference energy (mass) scale given by
\begin{align}
E_0 = \frac{4 M_1 M_2}{(2\pi)^{\frac{D-3}{2}} M_{Pl}},
\end{align}
where $M_{Pl}$ is the planck mass.
In this section we will plot $E_D(r)/E_0$ as a function of $M_{Pl}\, r$.  

\begin{figure}[htb!]
\centering
\includegraphics[width=8cm]{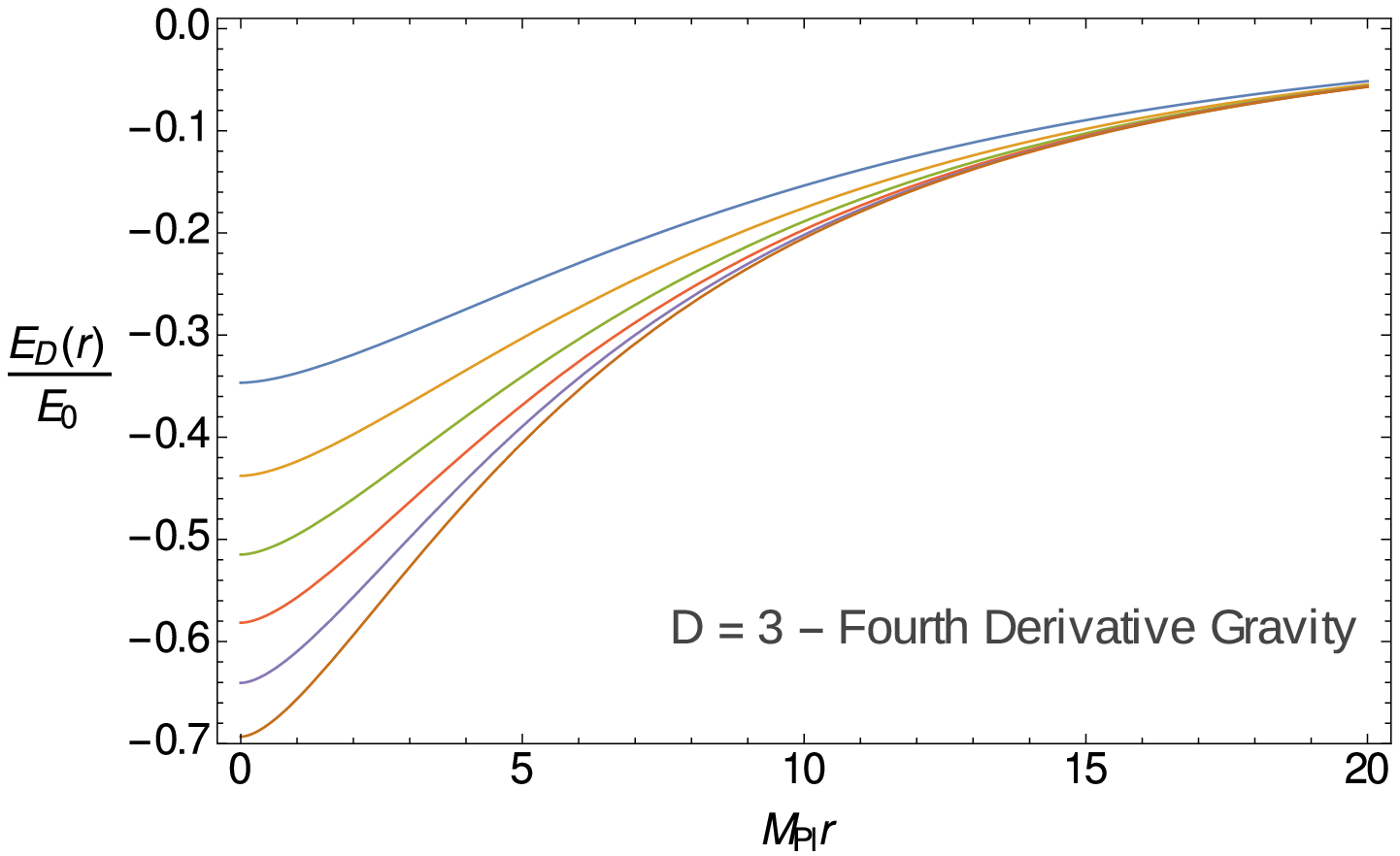} 
\includegraphics[width=8cm]{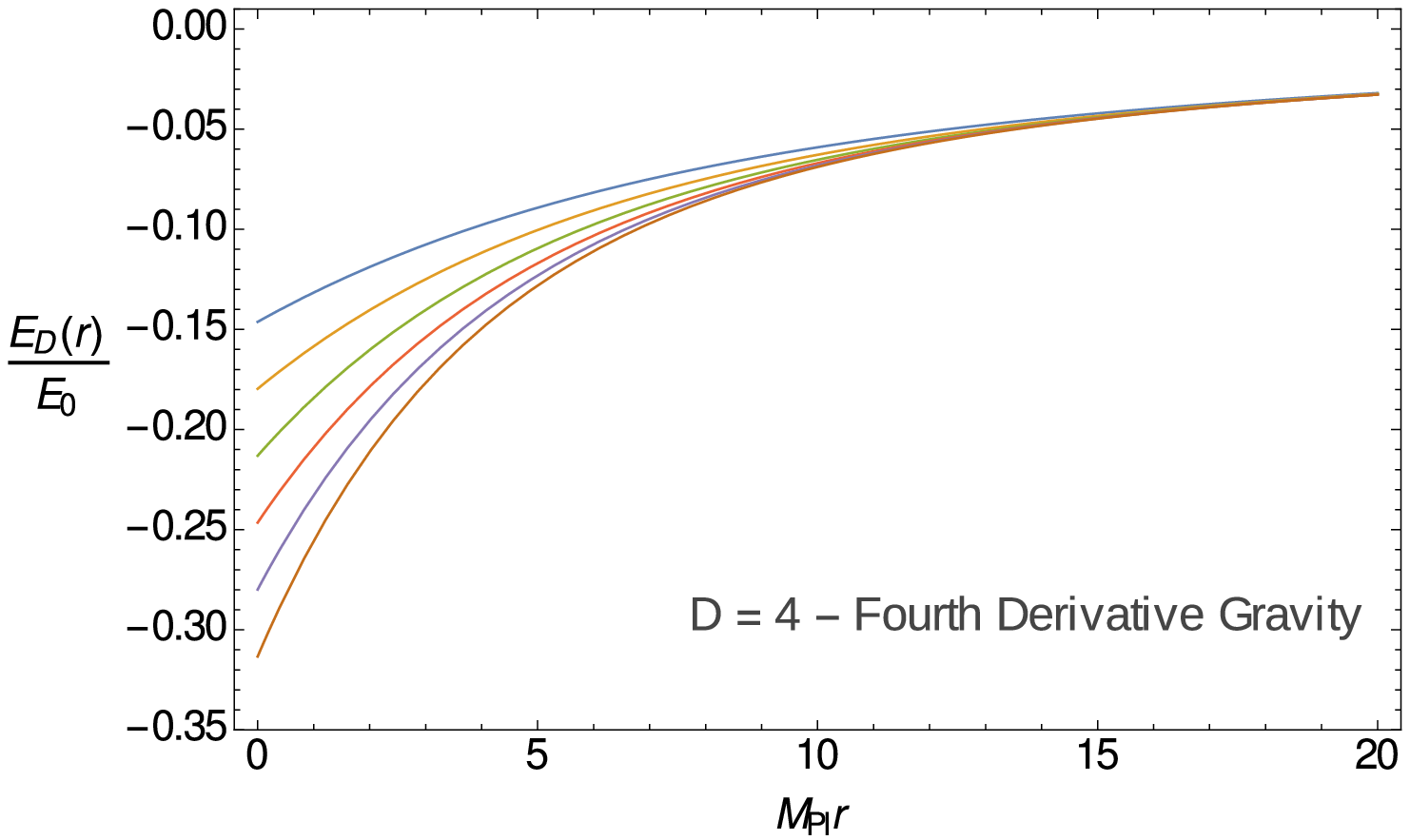} \\
\includegraphics[width=8cm]{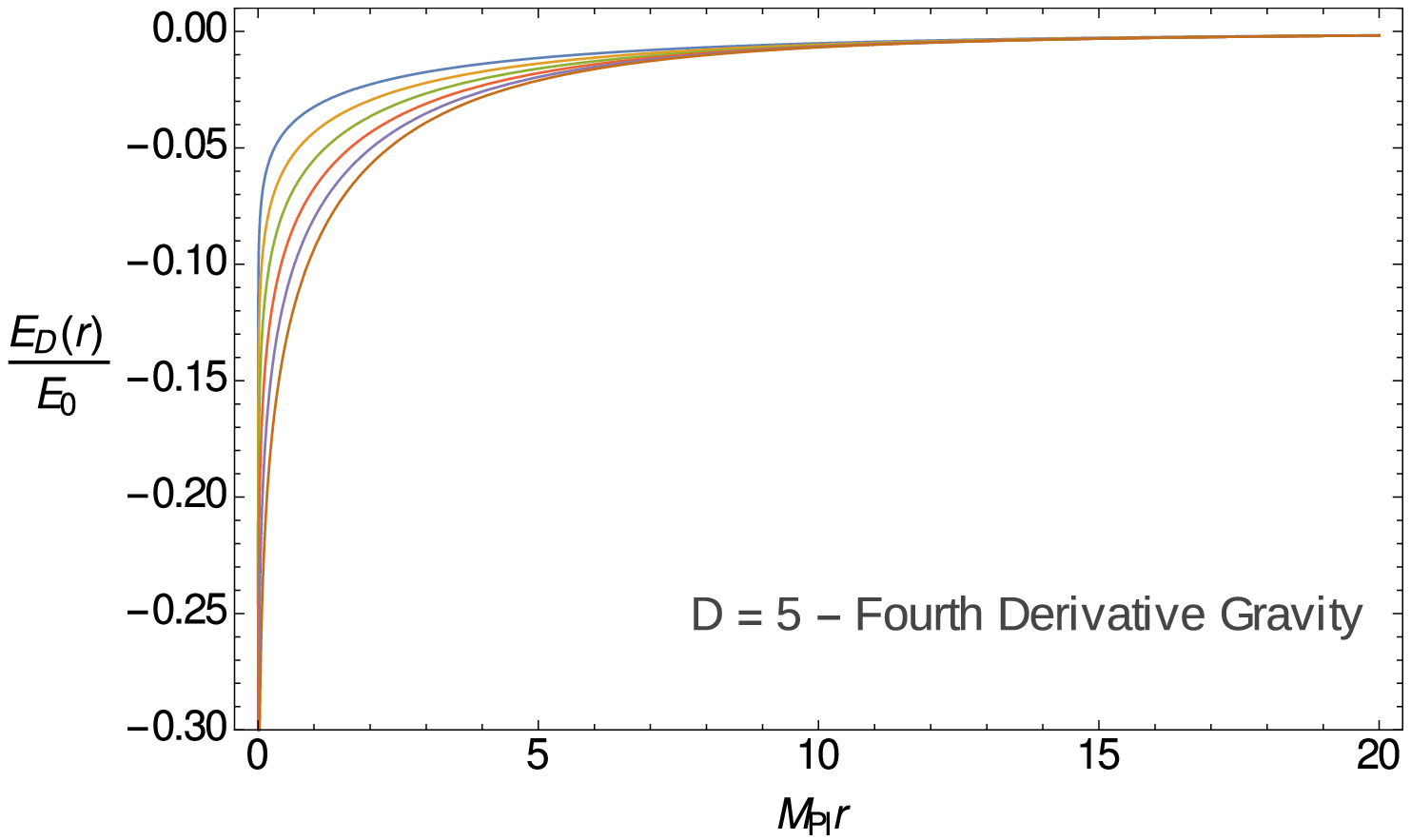}
\caption{\label{fig:i} Non-relativistic gravitational potential associated with fourth derivative models.}
\end{figure}

In Figure \ref{fig:i} we plot those results associated with fourth-derivative models for three different values of the space-time dimension ($D=3,\,4$ and $5$). In all cases we have considered several choices of the mass parameters $m_{(0),1}$ and $m_{(2),1}$ (close to the Planck mass $M_{Pl}$), however, their specific values are not relevant for this qualitative analysis. Also, it has been chosen $\sigma = 1$ in order to ensure the attractive behavior of the gravitational interaction. As one can see, both for $D=3$ and $D=4$ the non-relativistic potential energy turns out to be finite at $r=0$. On the other hand, the case $D=5$ exhibit the so-called Newtonian singularity at $r=0$. It  is important to emphasize that all these cases are in agreement with the aforementioned sufficient condition for the absence of Newtonian singularities. In addition, we remark that we are not plotting results for $D>5$, since their qualitative behavior are the same as $D=5$. 

\begin{figure}[htb!]
\centering 
\includegraphics[width=8cm]{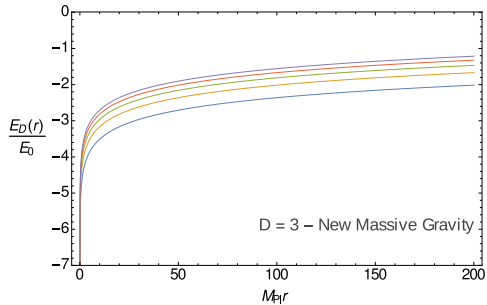}\qquad\qquad
\caption{\label{fig:ii} Non-relativistic gravitational potential - The case of ``New Massive Gravity''.}
\end{figure}

Now we turn our attention to the case of the so-called New Massive Gravity. As it was mentioned before, the NMG case is characterized by an special choice of the higher-derivative parameters such that $3\alpha_0 + 8 \beta_0 = 0$. In this case we have $m_{(0),1} \to \infty$, while $m_{(2),1}$ remains arbitrary. In addition, $\sigma = -1$ is required in order to keep the attractive behavior of the gravitational interaction. In Figure \ref{fig:ii} we plot those results associated with the case of NMG for several values of the mass parameter $m_{(2),1}$. As one can see, the NMG potential energy exhibit the so-called Newtonian singularity, as we could expect from the analytical results presented before.

\begin{figure}[htb!]
\centering 
\includegraphics[width=8cm]{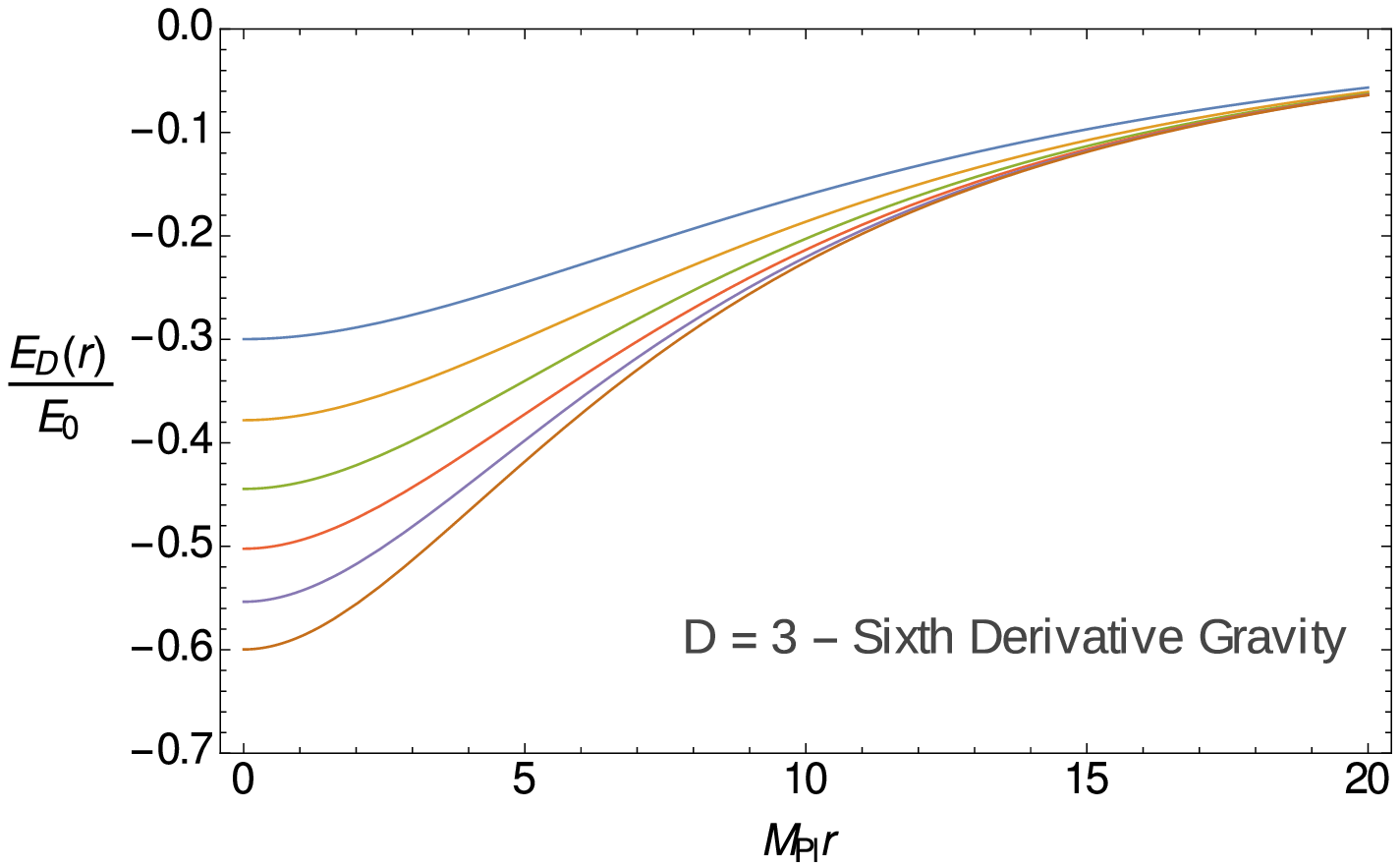} 
\includegraphics[width=8cm]{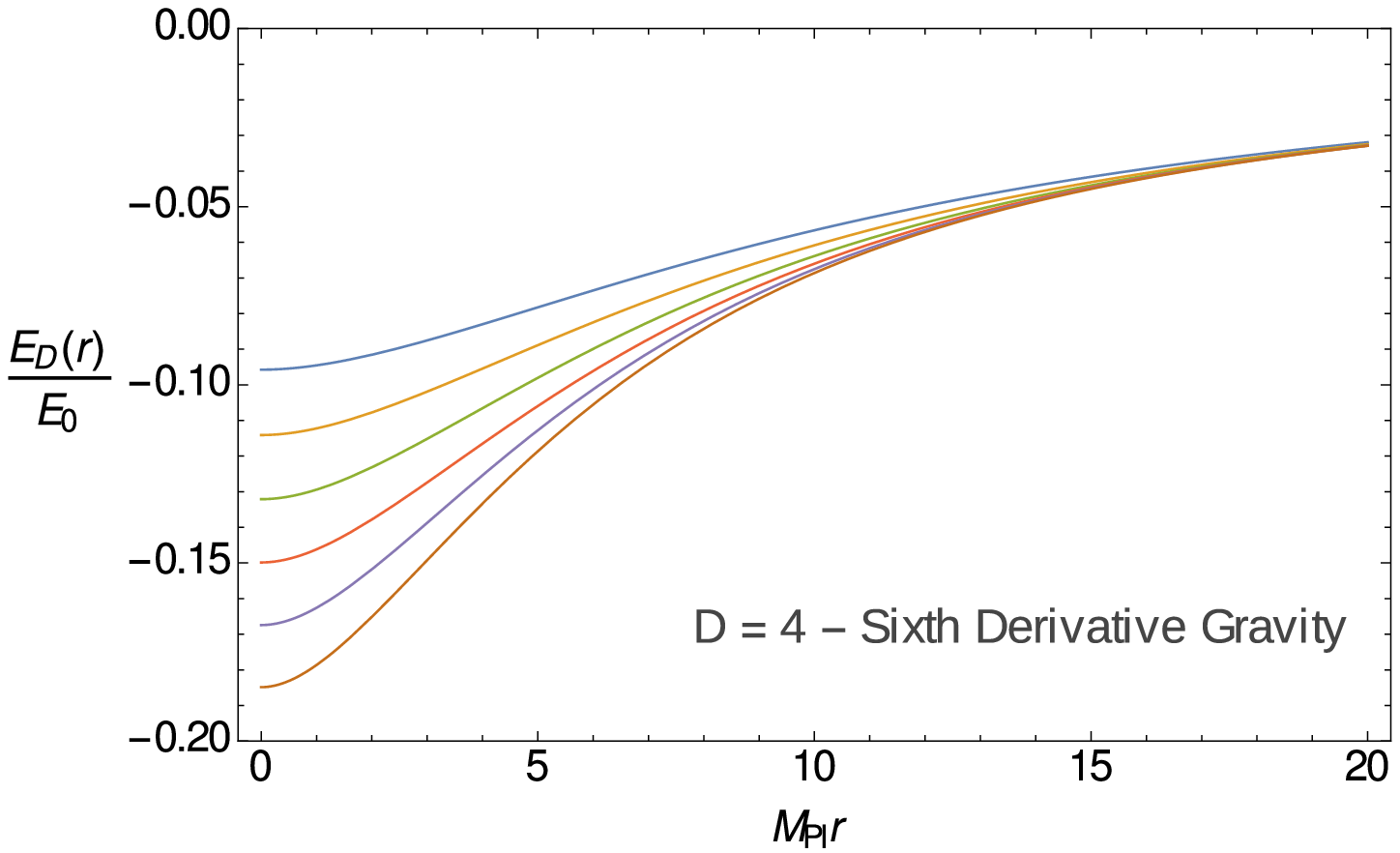} \\
\includegraphics[width=8cm]{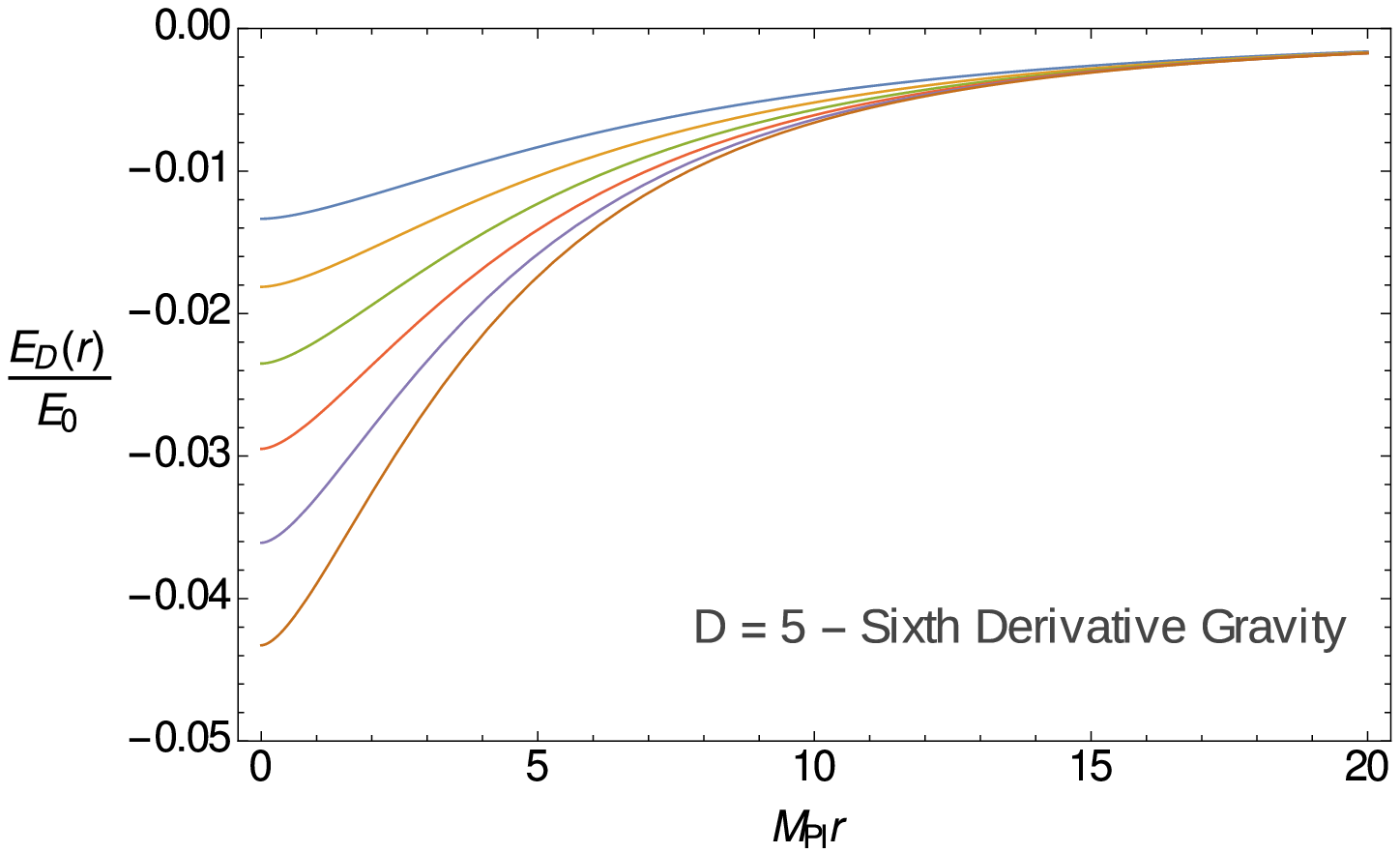} 
\includegraphics[width=8cm]{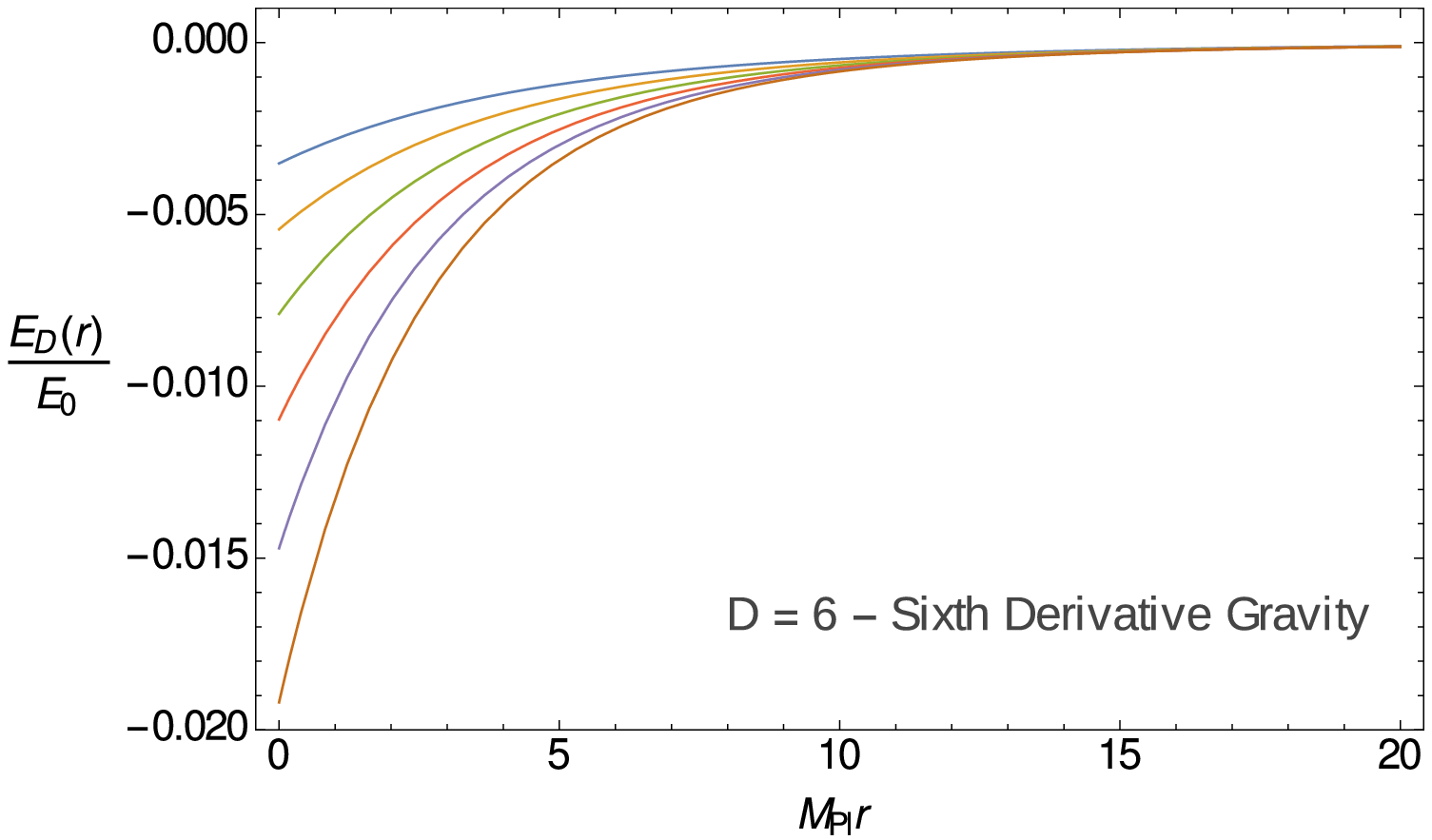} \\
\includegraphics[width=8cm]{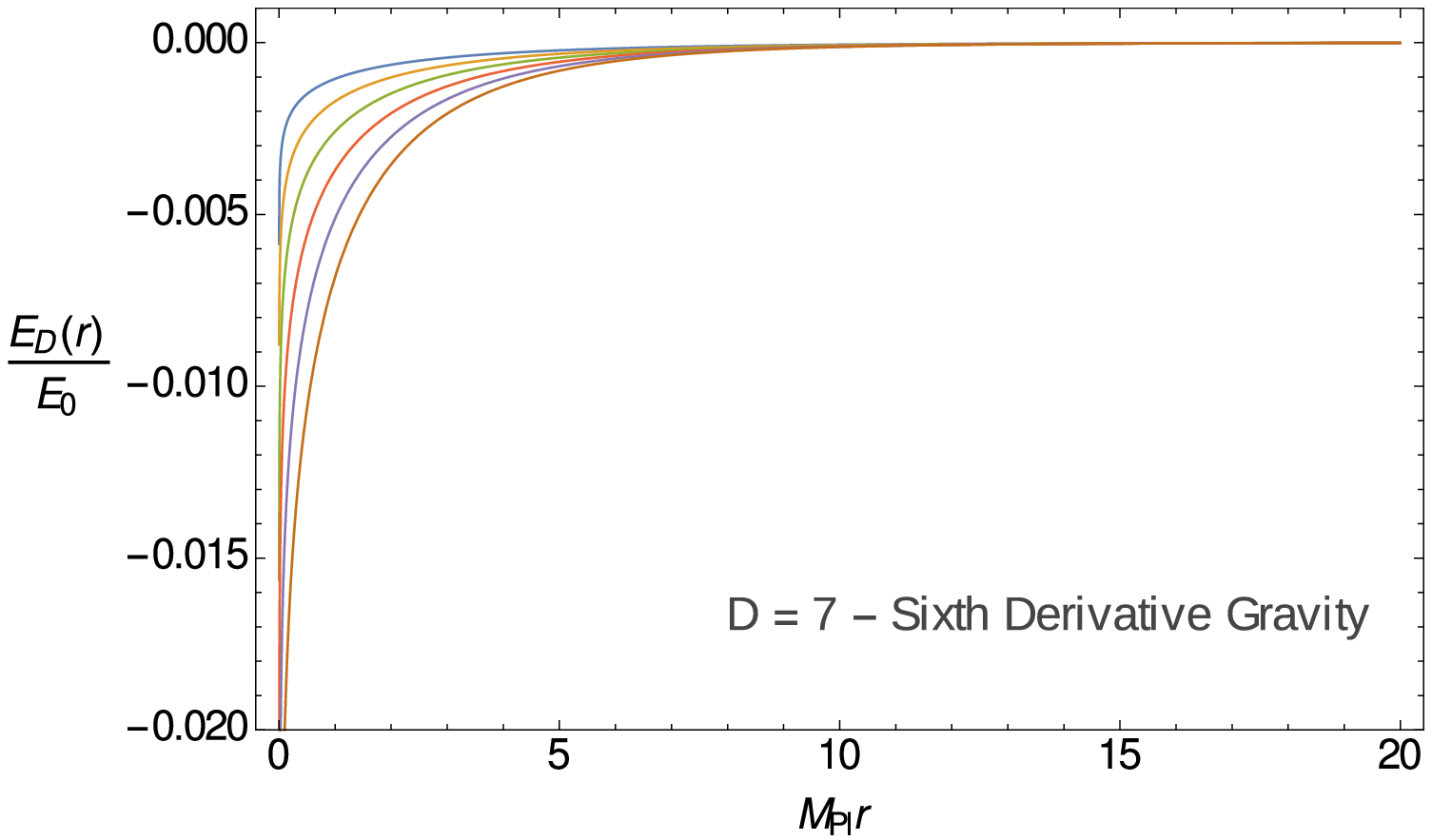}
\caption{\label{fig:iii} Non-relativistic gravitational potential associated with sixth derivative models.}
\end{figure}

Finally, in Figure \ref{fig:iii} we plot some results associated with sixth-derivative models in several space-time dimensions ($D=3,\,4,\,5,\,6$ and $7$). In all cases we have considered several choices for the massive parameters associated with higher-derivative terms, however, once again, their specific values are irrelevant for the general qualitative behavior discussed here. Remarkably, the non-relativistic potential corresponding to $D=3,\,4,\,5$ and $6$ turns out to be finite at $r=0$, while the case $D=7$ exhibit the so-called Newtonian singularity. Furthermore, the graphical behavior obtained with $D>7$ presents the same qualitative features than those corresponding to $D=7$. In all cases, those results presented are in agreement with sufficient condition for the cancellation of Newtonian singularities obtained before.

Although we have considered only the case of fourth- and sixth-derivative models, the analysis performed in this section is exhaustive and can be extended to higher-order gravities. In all cases the graphical behavior should reflect the conclusions obtained by analytical means in the previous section.

\section{UV Properties and Renormalizability \label{Renormalizability}}

As is well known, the motivation for considering  higher derivative theories of quantum gravity  as possible candidates for quantum gravity models is their good UV properties. As we have already mentioned, Stelle proved that fourth-derivative gravity theories are renormalizable (in $4$-dimension) to all orders in perturbation theory \cite{Stelle:1976gc}. Furthermore, theories containing sixth or higher derivatives may be formulated as superrenormalizable \cite{Shapiro:1997,Modesto:2012,Modesto:2012ys}. For completeness' sake in this section we shall analyze the UV properties of the general class of $D$-dimensional higher-derivative theories described by the action \eqref{action_general}.

The UV aspects of renormalizability of similar models have been studied in references \cite{Stelle:1976gc,Tomboulis,Shapiro:1997,Modesto:2012}. In addition, as it was stressed out in reference \cite{Mazumdar:UV}, power-counting criteria is just a hint for renormalizability and it should be checked with more involved methods, since there could be divergent sub-graphs. We note, however, that even if a power-counting renormalizable theory without Newtonian singularity turns out to be non-renormalizable, the conjecture discussed along this paper is still not falsified. The conjecture states that a finite potential at $r=0$ is a necessary condition for renormalizability, not a sufficient one.

Having said that, we proceed to construct the power-counting criteria for renormalizability of the general class of $D$-dimensional higher-derivative theories described by the action \eqref{action_general}. The propagators and vertices associated with this general class of higher-derivative theories have the following UV behavior\footnote{Recall that the parameters $q$ and $N$ were introduced in the previous section and are related to the number of derivatives contained in the action.}
\begin{eqnarray}
\textmd{Propagators} \sim \frac{1}{k^{2q+4}}
\qquad \textmd{and} \qquad 
\textmd{Vertices} \sim k^{2N+4}.
\end{eqnarray}
As a consequence, for an arbitrary Feynman diagram  the UV behavior of the loop integrations is given by
\begin{eqnarray}
\mathcal{I}_{UV}^{Loops} \sim \int (d^D k)^L \frac{(k^{2N+4})^V}{(k^{2q+4})^I} ,
\end{eqnarray}  
where $I$ is the number of internal lines, $L$ stands for the number of loops and $V$ denotes the   vertex-number. Therefore the superficial degree of divergence associated with  the integral may be cast as follows
\begin{eqnarray}
\delta = D L + (2 N + 4)V - (2 q + 4)I .
\end{eqnarray}
Now, bearing in mind the  topological relations
\begin{subequations}
	\begin{eqnarray}
	L-1 = I - V,
	\end{eqnarray}
	and
	\begin{eqnarray}
	2I + E = \sum_{n=3}^{\infty} n V_{n},
	\end{eqnarray}
\end{subequations}
where $E$ is the number of external lines and $V_{n}$ denotes the number of vertices connecting $n$-lines, we may recast the superficial degree of divergence as
\begin{eqnarray}\label{power-counting}
\delta = D   -   \sum_{n=3}^{\infty} \bigg[ \frac{n-2}{2}(2q+4 - D) - 2\lambda \bigg] V_n   + \bigg(\frac{2q+4-D}{2}\bigg)E  ,
\end{eqnarray}
where the parameter $\lambda$ is given by
\begin{eqnarray}
\lambda = \begin{cases}
p-q, &\textmd{if}\,\, q < p ,\\
0, &\textmd{if}\,\, q \geq p .
\end{cases}
\end{eqnarray}

As is well known,  the power-counting criteria for renormalizability requires that the superficial degree of divergence cannot depend on the number of vertices, therefore, we arrive at the following conditions for power-counting renormalizability
\begin{eqnarray}\label{criteria_powercounting}
2q+4 - D  = 0 \qquad \textmd{and} \qquad \lambda = 0 . 
\end{eqnarray}
The first condition relates the number of derivatives in the Ricci-squared sector to the dimension of spacetime. Consequently, in odd dimensions we cannot have power counting renormalizability  since in this case it would be  required   fractional powers of the d'Alembertian operator. The second condition implies essentially that $q \geq p $. This inequality tells us that the number of derivatives in the scalar curvature squared sector should not be greater than the number of derivatives in the Ricci squared sector.

Furthermore, the theory can also be formulated as power counting superrenormalizable. In fact, the condition for superrenormalizability demands that the superficial degree of divergence decreases with the number of vertices. Taking this condition into account, along with \eqref{power-counting}, we arrive at the  strong inequality 

\begin{eqnarray} 
2q+4 - D  \geq 4 \lambda .
\end{eqnarray}
As in the previous case, this  inequality relates the number of derivatives in the Ricci squared sector to the dimensionality of spacetime. In addition, if $p > q$ it determines a lower bound on the parameter $q$ in terms of the spacetime dimension and the number of derivatives in the scalar curvature squared sector.

It is important to call attention to the fact that the  discussion above relies upon  the assumption that the Lagrangian parameters, $\alpha$'s and $\beta$'s, are unrelated. In fact, if there existed  a link  between them, the power counting would be probably  slightly changed. To illustrate this, we discuss the so-called \textit{New Massive Gravity} (NMG) \cite{NMG}. This theory is characterized by the choice $F_1(\Box) = \alpha_0$ and $F_2(\Box) = \beta_0$ with the additional constraint 
\begin{eqnarray}
8\alpha_0 + 3 \beta_0 = 0.
\end{eqnarray}
In order to made the theory ghost-free we have to consider the ``wrong'' sign in the Einstein-Hilbert sector, \textit{i.e.} $\sigma = -1$, so that the NMG action becomes
\begin{eqnarray}
S_{\textmd{NMG}} = \int d^3 x \sqrt{|g|} \, \left( -\frac{2}{\kappa^2} R + \frac{\beta_0}{2\kappa^2} \left( R_{\mu\nu}^2 - \frac{3}{8}R^2 \right) \right). 
\end{eqnarray}
 At  first sigh, applying the power counting condition above,  NMG should be apparently  classified as superrenormalizable. If this was true, NMG  would be an example of a ghost-free and superrenormalizable model; however, this is not the case. Indeed,  the constraint $3 \alpha_0 + 8 \beta_0$ must be considered in such a way that the UV behavior of the tree-level propagator is given by $\sim 1/k^2$. Taking this into account, the correct power counting for the NMG is given by
\begin{eqnarray}
\delta_{\textmd{NMG}} = 3 - \frac{1}{2} E + \frac{1}{2} \sum_{n=3}^{\infty} (n+2) V_n .
\end{eqnarray}
As one can see, the superficial degree of divergence increases with the number of vertices and, as a consequence, the theory is power counting nonrenormalizable. Furthermore, it is important to emphasize that a complete proof of  the nonrenormalizability of the NMG was performed in reference \cite{Ohta}.

\section{Particle spectra and tree level unitarity}\label{unitarity_section}

One of the most intriguing problems concerning the formulation of quantum gravity considered as a  theory of fluctuations around the Minkowski spacetime is the incompatibility between unitarity and renormalizability. In what follows we shall investigate tree-level unitarity. As usual, this study can be done in terms of the pole structure of the saturated propagator given by
\begin{eqnarray}\label{Saturated_Propagator}
SP(k) = i \,T^{*\mu\nu}(k) \D_{\mu\nu,\alpha\beta}(k) T^{\alpha\beta}(k) ,
\end{eqnarray}
where $T^{\mu\nu}$ stands for an external conserved current. Using equation \eqref{Propagator}, we arrive at the  result
\begin{eqnarray}
SP(k) = \frac{i}{\sigma k^2 Q_2(k^2)}  \bigg( T^*_{\mu\nu}T^{\mu\nu} - \frac{1}{D-1} |T|^2 \bigg)-  \frac{i}{\sigma k^2 Q_0(k^2)} \frac{|T|^2}{(D-1)(D-2)} .
\end{eqnarray}
Considering that the form factors are given by \eqref{Polynomial_function}, by the factorization theorem for polynomials and partial fraction decomposion we arrive at the result
\begin{align}\label{Saturated_propagator}
&SP(k) = \frac{i}{\sigma k^2} \bigg( T^*_{\mu\nu}T^{\mu\nu} - \frac{1}{D-2} |T|^2 \bigg) +\nn \\
&+ \frac{|T|^2}{\sigma(D-1)(D-2)} \Bigg\{ \sum_{i=1}^{\tilde{N}+1} \frac{i\,\xi_{(0),i}}{k^2 - m_{(0),i}^2} + \sum_{i=1}^s \frac{i\,\zeta_{(0),i}}{k^2 - \eta_{(0),i}^2} + \sum_{i=1}^s \frac{i\,\zeta_{(0),i}^*}{k^2 - \eta_{(0),i}^{*\,2}}  \Bigg\}  +  \nn\\
&- \frac{1}{\sigma}\bigg( T^*_{\mu\nu}T^{\mu\nu} - \frac{1}{D-1} |T|^2 \bigg) \Bigg\{ \sum_{i=1}^{\tilde{q}+1} \frac{i\,\xi_{(2),i}}{k^2 - m_{(2),i}^2} + \sum_{i=1}^r \frac{i\,\zeta_{(2),i}}{k^2 - \eta_{(2),i}^2} + \sum_{i=1}^r \frac{i\,\zeta_{(2),i}^*}{k^2 - \eta_{(2),i}^{*\,2}}  \Bigg\}.
\end{align}
where the mass parameters are defined again by the roots of \eqref{Q_2} and \eqref{Q_0} and, additionally, we have defined
\begin{subequations}\label{definition_zetaxi}
	\begin{eqnarray}
	\zeta_{(0),i} = \prod_{\substack{j = 1\\ j \neq i}}^{s} \frac{\eta_{(0),j}^2}{\eta_{(0),j}^2 - \eta_{(0),i}^2} \prod_{j=1}^{s} \frac{\eta_{(0),j}^{*\,2}}{\eta_{(0),j}^{*\,2}  - \eta_{(0),i}^2} \prod_{j=1}^{\tilde{N}+1} \frac{m_{(0),j}^2}{m_{(0),j}^2 - \eta_{(0),i}^2},
	\end{eqnarray}
	\begin{eqnarray}
	\xi_{(0),i} = \prod_{j=1}^{s} \frac{\eta_{(0),j}^2}{\eta_{(0),j}^2 - m_{(0),i}^2} \prod_{j=1}^{s} \frac{\eta_{(0),j}^{*\,2}}{\eta_{(0),j}^{*\,2}  - m_{(0),i}^2} \prod_{\substack{j = 1\\ j \neq i}}^{\tilde{N}+1} \frac{m_{(0),j}^2}{m_{(0),j}^2 - m_{(0),i}^2},
	\end{eqnarray}
	\begin{eqnarray}
	\zeta_{(2),i} = \prod_{\substack{j = 1\\ j \neq i}}^{r} \frac{\eta_{(2),j}^2}{\eta_{(2),j}^2 - \eta_{(2),i}^2} \prod_{j=1}^{r} \frac{\eta_{(2),j}^{*\,2}}{\eta_{(2),j}^{*\,2}  - \eta_{(2),i}^2} \prod_{j=1}^{\tilde{q}+1} \frac{m_{(2),j}^2}{m_{(2),j}^2 - \eta_{(2),i}^2},
	\end{eqnarray}
	\begin{eqnarray}
	\xi_{(2),i} = \prod_{j=1}^{r} \frac{\eta_{(2),j}^2}{\eta_{(2),j}^2 - m_{(2),i}^2} \prod_{j=1}^{r} \frac{\eta_{(2),j}^{*\,2}}{\eta_{(2),j}^{*\,2}  - m_{(2),i}^2} \prod_{\substack{j = 1\\ j \neq i}}^{\tilde{q}+1} \frac{m_{(2),j}^2}{m_{(2),j}^2 - m_{(2),i}^2}.
	\end{eqnarray}
\end{subequations}
As usual, in order to investigate whether the theory exhibits ghosts in its spectrum, we compute the imaginary part of residues of the saturated propagator. Using equation \eqref{Saturated_propagator} we find 
\begin{subequations}
	\begin{eqnarray}
	\Im\Big[ \textmd{Res}\, SP(k^2 = 0) \Big] = \sigma^{-1}\,\bigg( T^*_{\mu\nu}T^{\mu\nu} - \frac{1}{D-2} |T|^2 \bigg)\bigg|_{k^2 = 0},
	\end{eqnarray}
	\begin{eqnarray}
	\Im\Big[ \textmd{Res}\, SP(k^2 = m_{(2),i}^2) \Big] = -\sigma^{-1}\, \xi_{(2),i} \bigg( T^*_{\mu\nu}T^{\mu\nu} - \frac{1}{D-1} |T|^2 \bigg)\bigg|_{k^2 = m_{(2),i}^2},
	\end{eqnarray}
	\begin{eqnarray}
	\Im\Big[ \textmd{Res}\, SP(k^2 = \eta_{(2),i}^2) \Big] = -\sigma^{-1} \,\Im(\zeta_{(2),i}) \bigg( T^*_{\mu\nu}T^{\mu\nu} - \frac{1}{D-1} |T|^2 \bigg)\bigg|_{k^2 = \eta_{(2),i}^2},
	\end{eqnarray}
	\begin{eqnarray}
	\Im\Big[ \textmd{Res}\, SP(k^2 = \eta_{(2),i}^{*\,2}) \Big] = -\sigma^{-1} \,\Im(\zeta^*_{(2),i}) \bigg( T^*_{\mu\nu}T^{\mu\nu} - \frac{1}{D-1} |T|^2 \bigg)\bigg|_{k^2 = \eta_{(2),i}^{*\,2}},
	\end{eqnarray}
	\begin{eqnarray}
	\Im\Big[ \textmd{Res}\, SP(k^2 = m_{(0),i}^2) \Big] =  \frac{\sigma^{-1} \,\xi_{(0),i}}{(D-1)(D-2)} |T|^2 \Big|_{k^2 = m_{(0),i}^2},
	\end{eqnarray}
	\begin{eqnarray}
	\Im\Big[ \textmd{Res}\, SP(k^2 = \eta_{(0),i}^2) \Big] =  \frac{\sigma^{-1} \,\Im(\zeta_{(0),i})}{(D-1)(D-2)} |T|^2 \Big|_{k^2 = \eta_{(0),i}^2},
	\end{eqnarray}
	\begin{eqnarray}
	\Im\Big[ \textmd{Res}\, SP(k^2 = \eta_{(0),i}^{*\,2}) \Big] =  \frac{\sigma^{-1}\, \Im(\zeta_{(0),i}^*)}{(D-1)(D-2)} |T|^2 \Big|_{k^2 = \eta_{(0),i}^{*\,2}}.
	\end{eqnarray}
\end{subequations}
Assuming that the real masses obey the  hierarchy\footnote{It is important to emphasize that this ordering can always be achieved by relabeling the masses.}
\begin{eqnarray}
m_{(2),1}^2 < m_{(2),2}^2 < \cdots < m_{(2),\tilde{q}+1}^2 \qquad \textmd{and} \qquad m_{(0),1}^2 < m_{(0),2}^2 < \cdots < m_{(0),\tilde{N}+1}^2,
\end{eqnarray}
we arrive at the  conclusion
\begin{eqnarray}
\begin{cases}
\xi_{(2),i} > 0, \quad \textmd{if $i=$ odd}, \\
\xi_{(2),i} < 0, \quad \textmd{if $i=$ even},
\end{cases}
\qquad \textmd{and} \qquad
\begin{cases}
\xi_{(0),i} > 0, \quad \textmd{if $i=$ odd}, \\
\xi_{(0),i} < 0, \quad \textmd{if $i=$ even}.
\end{cases}
\end{eqnarray}
From now one we divide our 
analysis  in two cases: $D\geq 4$ and $D =3$.

\subsection{Case I - $D \geq 4$:}

We first consider the case where  $D\geq 4$. Taking into account the set of inequalities (valid for $D \geq 4$)
\begin{subequations}
	\begin{eqnarray}\label{energia_momentum_unitariedade}
	\bigg( T^*_{\mu\nu}T^{\mu\nu} - \frac{1}{D-2} |T|^2 \bigg)\bigg|_{k^2 = 0} > 0,
	\end{eqnarray}
	and
	\begin{eqnarray}
	\bigg( T^*_{\mu\nu}T^{\mu\nu} - \frac{1}{D-1} |T|^2 \bigg)\bigg|_{k^2 = \mu^2} > 0, \qquad \mu^2 = m_{(2),i}^2,\eta_{(2),i}^{2},\eta_{(2),i}^{*\,2}
	\end{eqnarray}
\end{subequations}
we obtain the  results
\begin{subequations}\label{Residues}
	\begin{eqnarray}\label{Residue_graviton}
	\Im\Big[ \textmd{Res}\, SP(k^2 = 0) \Big] \,\,\begin{cases}
	> 0 , \quad \textmd{for $\sigma = +1$}\\
	< 0 , \quad \textmd{for $\sigma = -1$}
	\end{cases} ,
	\end{eqnarray}
	\begin{eqnarray}\label{residue_m_2_odd}
	\Im\Big[ \textmd{Res}\, SP(k^2 = m_{(2),i}^2) \Big] \,\,\begin{cases}
	< 0 , \quad \textmd{for $\sigma = +1$}\\
	> 0 , \quad \textmd{for $\sigma = -1$}
	\end{cases} ,\quad \textmd{if $i=$ odd} ,
	\end{eqnarray}
	\begin{eqnarray}\label{residue_m_2_even}
	\Im\Big[ \textmd{Res}\, SP(k^2 = m_{(2),i}^2) \Big] \,\,\begin{cases}
	> 0 , \quad \textmd{for $\sigma = +1$}\\
	< 0 , \quad \textmd{for $\sigma = -1$}
	\end{cases} , \quad \textmd{if $i=$ even} ,
	\end{eqnarray}
	\begin{eqnarray}\label{residue_m_0_odd}
	\Im\Big[ \textmd{Res}\, SP(k^2 = m_{(0),i}^2) \Big] \,\,\begin{cases}
	> 0 , \quad \textmd{for $\sigma = +1$}\\
	< 0 , \quad \textmd{for $\sigma = -1$}
	\end{cases} , \quad \textmd{if $i=$ odd} ,
	\end{eqnarray}
	\begin{eqnarray}\label{residue_m_0_even}
	\Im\Big[ \textmd{Res}\, SP(k^2 = m_{(0),i}^2) \Big] \,\,\begin{cases}
	< 0 , \quad \textmd{for $\sigma = +1$}\\
	> 0 , \quad \textmd{for $\sigma = -1$}
	\end{cases} , \quad \textmd{if $i=$ even} .
	\end{eqnarray}
\end{subequations}
In addition, considering $\Im(\zeta_{(I),i}) = -\Im(\zeta^*_{(I),i})$ (where $I = 0,2$), we may conclude that:
\begin{subequations}
	\begin{eqnarray}\label{residue_complex_1}
	\textmd{if} \quad \Im\Big[ \textmd{Res}\, SP(k^2 = \eta_{(I),i}^2) \Big] > 0   \quad \Rightarrow \quad \Im\Big[ \textmd{Res}\, SP(k^2 = \eta_{(I),i}^{*\,2}) \Big] < 0,
	\end{eqnarray}
	and
	\begin{eqnarray}\label{residue_complex_2}
	\textmd{if} \quad \Im\Big[ \textmd{Res}\, SP(k^2 = \eta_{(I),i}^2) \Big] < 0   \quad \Rightarrow \quad \Im\Big[ \textmd{Res}\, SP(k^2 = \eta_{(I),i}^{*\,2}) \Big] > 0.
	\end{eqnarray}
\end{subequations} 
Taking into account the set of inequalities above we may list some conclusions:
\begin{itemize}
	\item Although our former discussion applies for  both  $\sigma = +1$ and $\sigma  = -1$, relation \eqref{Residue_graviton} imply $\sigma = +1$, since we expect a physical massless spin-2 particle in the spectrum of the theory corresponding to the usual graviton.
	\item The usual drama of higher derivative theories persists as long as real poles are present. In this case, the alternating signs of the parameters $\xi_{(2),i}$ and $\xi_{(0),i}$ ensure the existence of at least one ghost-like particle. As usual the existence of such particles in the spectrum may lead to a nonunitary $S$-matrix in the context of perturbation theory.
	\item Since the complex poles always appear in pairs, one of them being the complex conjugate of the other, we may conclude that for each complex ``physical'' particle (not a ghost-like state) there will be a complex ghost corresponding to the complex conjugated of the former. Although complex ghosts may appear in the particle spectrum of the theory they may not cause problems with the unitarity of the $S$-matrix. In fact, higher- derivative gravity systems with complex ghosts have been recently studied by Modesto and Shapiro \cite{Modesto:2016,Shapiro:2016} and there is hope that this kind of theories may be formulated as unitary in the Lee-Wick sense.
\end{itemize}

The situation now is clear: as long as higher derivatives are implemented by means of polynomial functions of the d'Alembertian operator the presence of at least one massive ghost-like particle appears to be unavoidable. In fact, there are only three cases of higher-derivative gravity models, constructed with polynomial functions like \eqref{Polynomial_function}, where the particle spectrum do not exhibits a massive ghost like state. The first one occurs with the choice $F_1(\Box) = \alpha_0$ and $F_2(\Box) = 0$ - in this case the propagator has only two poles, $k^2 = 0$ and $k^2 = m_{(0),1}^2$ and we conclude that both poles corresponds to physical particles, \textit{i.e.} the theory is ghost-free (at least in tree-level analysis). We remark that, as  expected, this theory is classified as non-renormalizable by the power-counting criteria of section \ref{Renormalizability}. The the other two cases occur in $D=3$ and will be considered in the next section.

\subsection{Case II - $D = 3$:}

Now we consider the 3-dimensional case. In this situation the energy-momentum tensor satisfies the relations
\begin{subequations}
	\begin{eqnarray}\label{energia_momentum_unitariedade_3D}
	\bigg( T^*_{\mu\nu}T^{\mu\nu} - |T|^2 \bigg)\bigg|_{k^2 = 0} = 0,
	\end{eqnarray}
	and
	\begin{eqnarray}\label{energia_momentum_unitariedade_3D_2}
	\bigg( T^*_{\mu\nu}T^{\mu\nu} - \frac{1}{2} |T|^2 \bigg)\bigg|_{k^2 = \mu^2} > 0, \qquad \mu^2 = m_{(2),i}^2,\eta_{(2),i}^{2},\eta_{(2),i}^{*\,2}.
	\end{eqnarray}
\end{subequations}
First of all, equation \eqref{energia_momentum_unitariedade_3D} implies in the result
\begin{eqnarray}\label{Residue_graviton_3D}
\Im\Big[ \textmd{Res}\, SP(k^2 = 0) \Big] =0,
\end{eqnarray}
while  (3,20b) tells us that there is no massless propagating mode in the 3-dimensional theory. 
As a consequence of the last equation we cannot fix the parameter $\sigma$ as we have done in the $D \geq 4$ case. As far as the massive poles (real and complex) are concerned,  the inequality \eqref{energia_momentum_unitariedade_3D_2} implies that the previous constraints,  namely \eqref{residue_m_2_odd}-\eqref{residue_complex_2}, remain valid as well  as the comments about massive ghosts in 3D.

As it was mentioned previously, in a 3-dimensional spacetime there are two ghost-free higher-derivative theories. The first is obtained by choosing the coefficients in such a way that $F_1(\Box) = \alpha_0$ and $F_2(\Box) = 0$. Consequently, the only particle in the spectrum corresponds to the pole $k^2 = m_{(0),1}^2$ and the theory is ghost-free, since it can be verified that the residue at this pole is positive.

The other one is the previously discussed \textit{New Massive Gravity} (NMG) \cite{NMG}. In this theory the free propagator exhibits two simple poles at $k^2 = 0$ and $k^2 = m_{(2),1}^2 \equiv 4/\beta_0$. Contrary to  what happens in dimensions other than 3, the massless pole do not propagate as a physical particle, since $  \textmd{Res}\, SP(k^2 = 0)|_{D=3} = 0$. Usually the second pole $k^2 = m_{(2),1}^2$ would propagate as a massive ghost, but here the ``wrong'' sign of the Einstein-Hilbert terms leads to a positive valued residue:
\begin{eqnarray}
\Im\Big[ \textmd{Res}\, SP(k^2 = m_{(2),i}^2) \Big]\Big|_{\textmd{NMG}} = \bigg( T^*_{\mu\nu}T^{\mu\nu} - \frac{1}{2} |T|^2 \bigg)\bigg|_{k^2 = m_{(2),1}^2} > 0.
\end{eqnarray}
Consequently the spectrum of  NMG  has only a single massive physical particle with spin-$2$ and there is no ghost-like state in the tree- level propagator. 

Last but not least, we would like to draw the reader's attention to reference \cite{Ohta_unitarity}, where an interesting study of unitarity higher derivative theories was performed on flat and maximally symmetric spaces.

\section{Relating renormalizability, unitarity and potential energy \label{Conjecture}}

In his seminal paper about renormalizability of higher- derivative quantum gravity \cite{Stelle:1976gc}, Stelle  hinted at the possibility of an  interesting connection between renormalizability of higher- derivatives theories of quantum gravity and the cancellation of Newtonian singularities in the  interparticle potential energy. It was later proposed that this connection is a general property of $D$-dimensional higher derivative theories of quantum gravity \cite{Accioly:2013hwa}. Essentially this surmise states that a renormalizable theory of quantum gravity should not present the so-called Newtonian singularity. Recently this conjecture was probed in the case of fourth- and sixth- derivative theories of gravity \cite{Accioly:2017-1,Accioly:2017-2}. We are now ready to verify this conjecture for a general class of $D$-dimensional higher- derivative theories of quantum gravity described by the action \eqref{action_general} with form factors given by \eqref{Polynomial_function}. 

Let us recall two important results from our last two sections. First of all, investigating the UV properties of the general class of higher -derivative theories under consideration, we found the following necessary conditions for (super)renormalizability:
\begin{itemize}
	\item $ 2q+4-D = 0$  $\,\sim\,$  power counting renormalizability;
	\item  $ 2q+4-D \geq 4\lambda$  $\,\sim\,$  power counting super-renormalizability,
\end{itemize}
where we remind that, for the convenience of the reader, $D$ denotes the dimension of space-time, the parameter $\lambda$ is defined to be
\begin{eqnarray}
\lambda = \begin{cases}
p-q, &\textmd{if}\,\, q < p ,\\
0, &\textmd{if}\,\, q \geq p ,
\end{cases}
\end{eqnarray}
and $q$ and $p$ denote, respectively, the degree of the polynomial form factor of the Ricci tensor sector and the degree of the polynomial form factor of the Ricci scalar sector (as given by \eqref{Polynomial_function}).

Furthermore, after an exhaustive investigation on the behavior of the interparticle potential energy for small distances, we found the sufficient conditions for the cancellation of Newtonian singularities:
\begin{itemize}
	\item $2q + 4 - D \geq 0$, for even dimensions;
	\item $2q + 3 - D \geq 0$, for odd dimensions.\\
\end{itemize}
Putting all these informations together, it is not difficult to conclude that the necessary conditions for (super)renormalizability automatically implies in the  the sufficient condition for the cancellation of Newtonian singularities. Summing up:
\begin{eqnarray}
\textmd{Power counting (super)renormalizability} \quad \Rightarrow \quad \textmd{Finite potential energy at $r=0$}. \nn
\end{eqnarray}
This completes our examination of the aforementioned conjecture for the general class of theories under consideration. It should be emphasized that inverse of this conjecture is not necessarily true, \textit{i.e.} the cancellation of Newtonian singularities does not implies renormalizability. In fact, as it was pointed out by Giacchini, it  is not difficult to construct an example of a higher- derivative model with finite potential energy at $r=0$ and power counting nonrenormalizable \cite{Breno:2016}.

Finally, let us discuss in passing  the role of (non)unitarity on the connection between renormalizability and Newtonian singularities. As we can see from the previous section, ghost-free higher- derivative (local) theories are not compatible with renormalizability. The reason for this point relies on the crucial role played by the ghost-like particles in the improvement of the tree-level propagator. Furthermore, ghost  are also necessary in the cancellation of Newtonian singularities \cite{Tiberio:2015,Breno:2016}. For instance, is not difficult to verify that those higher- derivative theories which are ghost-free, have a divergent potential energy at $r=0$. 

Indeed, the fact that ghost-like particles are necessary for the cancellation of  Newtonian singularities has an interesting explanation. From the classical point of view, ghost-like particles correspond to negative-energy propagation modes (which give rises to Ostrogradsky instabilities, for instance). Taking into account, from an heuristic point of view, that the interparticle potential energy is given by the sum of  individual  energies  associated with each propagation mode, it is necessary to have parts with opposite signs in order to have some kind of cancellation. Therefore, negative-energy propagation modes are necessary for the cancellation of Newtonian singularities. 

The role of ghost-free particles in the mechanism for  cancellation of Newtonian singularities was recently explored in the literature. In fact, Modesto and collaborators demonstrated that $4$-dimensional theories described by \eqref{action_general}, with $F_1(\Box)$ and $F_2(\Box)$ being polynomial functions  with the same degree, \textit{i.e.}  the same number of ghosts and physical particles, the cancellation of Newtonian singularities occurs \cite{Tiberio:2015}. Later, Giacchini demonstrated the equal number of ghosts and physical particle is not a necessary condition for the cancellation of Newtonian singularities in $4$-dimensional higher- derivative gravity \cite{Breno:2016}. The  necessary condition is that the particle spectra of the theory should contain \textit{at least} a massive ghost and a massive physical particle (besides the usual massless graviton). In the case of $3$-dimensional theories, it is not difficult to adapt Giacchini's demonstration in order to get the same conclusion. However, in the case of spacetime with dimension higher than four the situation is more subtle. Although we have no demonstration, the above proposition appears to be valid as well, nevertheless, the minimal number of both massive ghosts and physical particles increases with the dimension of  spacetime. 

\section{Final remarks \label{conclusion}}

In this paper we discussed some aspects concerning the tree-level unitarity, renormalizability properties and their relation to the classical context. Our primary aim in this paper was to prove a conjecture which states that (super)renormalizable theories of quantum gravity do not present the so-called Newtonian singularity in the potential energy. As we have seen, the necessary condition for power counting (super)renormalizability automatically leads to  a sufficient condition for the cancellation of Newtonian singularities. We analyzed  the role of ghost-like particles in the mechanism of cancellation of Newtonian singularities as well.

We also performed an exhaustive research of the particle spectra and the presence of ghost-like particles and showed that the appearance of them  in higher- derivative theories seems to be unavoidable. However, there are two exceptions to this statement : a special case of $f(R)$-theories and  New Massive Gravity. 

We studied also the UV properties of a general class of higher- derivative theories. We found the necessary conditions for power-counting (super)renormalizability, relating the number of derivatives to spacetime dimension. Unfortunately, the models which are tree-level ghost-free turns out to be nonrenormalizable. This fact clearly shows  the impossibility of reconciling renormalizability with unitarity in quantum gravity. A possible way to circumvent  this problem is to use  nonlocal theories of quantum gravity \cite{Modesto:2014,Modesto:2017} or to utilize the  Lee-Wick formulation of unitarity \cite{Modesto:2016,Shapiro:2016}. In two recent papers, Anselmi and Piva developed a new formulation of Lee-Wick theories \cite{Anselmi-1,Anselmi-2} and it certainly deserves some investigation in the context of higher- derivative quantum gravities. 

It is worth mentioning that we have not yet found  a complete proof of the necessary condition for the cancellation of Newtonian singularities in arbitrary dimensions. Our calculations were made  with the help of an algebraic manipulation system;  on the other hand,  their analytical consequences open a new road for finding a general solution to this issue.  

Finally, we call attention to the fact that our  discussion regarding the connection between (super)renormalizability and the cancellation of Newtonian singularities are based on the assumption of polynomial (local) form factors. We hope that our conclusions may be generalized for nonlocal theories; however, the mechanism concerning the  cancellation of Newtonian singularities depends on the specific form of the functions $F_1(\Box)$ and $F_2(\Box)$.

\section*{Acknowledgments}

The authors are very grateful to \textit{Conselho Nacional de Desenvolvimento Cient\'ifico e Tecnol\'{o}gico} (CNPq) and \textit{Funda\c{c}\~ao de Amparo \`a Pesquisa do Estado do Rio de Janeiro} (FAPERJ) for their financial support. G.P.B. is thankful to The Abdus Salam International Centre for Theorerical Physics for the hospitality when part of this work was done.



\begin{thebibliography}{99}
	
	
	\bibitem{Kiefer} 
	Kiefer, C., \emph{Quantum Gravity}, Oxford University Press, New York (2004).
	
	\bibitem{Oriti} 
	Oriti, D., \emph{Approaches to quantum gravity: toward a new understanding of space, time and matter}, 	Cambridge University Press, New York (2009).
	
	\bibitem{Shapiro} 
	Buchbinder, I.L., Odintsov, S.D. and Shapiro, I.L. \emph{Effective Action in Quantum Gravity}, IOP Publishing, Bristol (1992).
	
	\bibitem{Donoghue}
	Donoghue, J.F., \emph{General relativity as an effective field theory: The leading quantum corrections}, \emph{Phys. Rev. D} \textbf{50} (1994) 3874.
	
    \bibitem{Percacci} 
	Percacci, R., \emph{An Introduction to Covariant Quantum Gravity and Asymptotic Safety}, World Scientific Pub Co. Pte. Ltd., Singapore (2017).
    
    \bibitem{Stelle:1976gc}
	Stelle, K. S., \emph{Renormalization of Higher-Derivative Quantum Gravity}, \emph{Phys. Rev. D} {\bf16} (1977) 953.
	
	\bibitem{Tomboulis}
	Tomboulis, E.T., \textit{Unitarity in Higher Derivative Quantum Gravity}, \textit{Phys.Rev.Lett.} 52 (1984) 1173.
	
	\bibitem{Shapiro:1997}
	Asorey, M., López, J.L. and Shapiro, I.L., \emph{Some remarks on high derivative quantum gravity}, \emph{Int. J. Mod. Phys. A} {\bf 12} (1997) 5711.
	
	\bibitem{Modesto:2012}
	Modesto, L., \emph{Super-renormalizable quantum gravity}, \emph{Phys. Rev. D} {\bf 86} (2012) 044005.
	
	\bibitem{Modesto:2012ys}
	Modesto, L., \emph{Super-renormalizable Multidimensional Quantum Gravity: Theory and Applications}, \emph{Astron. Rev.} {\bf 8} (2013) 4.
	
	\bibitem{Modesto:2016}
	Modesto, L., \emph{Super-renormalizable or finite Lee-Wick quantum gravity}, \emph{Nucl. Phys. B} {\bf 909} (2016) 584.
	
	\bibitem{Shapiro:2016}
	Modesto, L. and Shapiro, I.L., \emph{Superrenormalizable quantum gravity with complex ghosts}, \emph{Phys. Lett. B} {\bf 755} (2016) 279.
	
	\bibitem{Yamamoto:1969}
	Yamamoto, H., \emph{Convergent Field Theory with Complex Masses}, \emph{Prog. Theor. Phys.} {\bf 42} (1969) 707.
	
	\bibitem{Biswas:2012}
	Biswas, T., Gerwick, E., Koivisto, T. and Mazumdar, A., \emph{Towards Singularity- and Ghost-Free Theories of Gravity}, \emph{Phys. Rev. Lett} {\bf 108} (2012) 031101.
	
	\bibitem{Modesto:2014}
	Modesto, L. and Rachwal, L., \emph{Super-renormalizable and finite gravitational theories}, \emph{Nucl. Phys. B} {\bf 889} (2014) 228.
	
	\bibitem{Modesto:2017}
	Modesto, L. and Rachwal, L., \emph{Nonlocal quantum gravity: A review}, \emph{Int. J. Mod. Phys. D} {\bf 26} (2017) 1730020.
	
	\bibitem{Shapiro:2015}
	Shapiro, I.L., \textit{Counting ghosts in the ``ghost-free'' non-local gravity}, \emph{Phys. Lett. B}, \textbf{744} (2015) 67.
	
	\bibitem{Biswas:2005}
	Biswas, T., Mazumdar, A. and Siegel, W., \textit{Biswas, T., Gerwick, E., Koivisto, T. and Mazumdar, A.},  \textit{JCAP} \textbf{0603} (2006) 009.
	
	\bibitem{Koshelev:2017}
	Koshelev, A.S. and Mazumdar, A., \textit{Do massive compact objects without event horizon exist in infinite derivative gravity?}, Phys. Rev. D 96 (2017) 084069.
	
	\bibitem{Talaganis:2016}
	 Talaganis, S. and Mazumdar, A., \textit{High-energy scatterings in infinite-derivative field theory and ghost-free gravity}, \textit{Class. Quantum Gravity} \textbf{33} (2016) 145005.
	
	\bibitem{Accioly:2013hwa}
	Accioly, A., Helayel-Neto, J., Scatena, E. and Turcati, R., \emph{Solving the riddle of the incompatibility between renormalizability and unitarity in $N$-dimensional Einstein gravity enlarged by curvature-squared terms}, \emph{Int. J. Mod. Phys. D} {\bf 22} (2013) 1342015.
	
	\bibitem{Accioly:2017-1}
	Accioly, A., Almeida, J., Brito, G.P. and Correia, G. , \emph{Renormalizability in D-dimensional higher-order gravity}, \textit{Phys.Rev. D} \textbf{95} (2017) 084007.
	
	\bibitem{Accioly:2017-2}
	Accioly, A.,Correia, G., Brito, G.P., Almeida, J. and Herdy, W., \emph{Relating renormalizability of D-dimensional higher-order electromagnetic and gravitational models to the classical potential at the origin}, \textit{Mod. Phys. Lett. A} \textbf{32} (2017) 1750048.
    
    \bibitem{Scaleinvariant:2017}
	Myung, Y.S., \textit{Renormalizability and Newtonian potential in scale-invariant gravity}, arXiv:1708.03451 [gr-qc].
	
    \bibitem{Breno:2016}
	Giacchini, B.L., \emph{On the cancellation of Newtonian singularities in higher-derivative gravity}, \textit{Phys.Lett. B} \textbf{766} (2017) 306.
	
	\bibitem{Tiberio:2015}
	Modesto, L., Paula Netto, T. and Shapiro, I.L., \emph{On Newtonian singularities in higher derivative gravity models}, \emph{Journ. High Energy Phys.} {\bf 04} (2015) 098.
	 	 
	 \bibitem{Biswas:2013}
	 Biswas, T.,  Koivisto, T. and Mazumdar, A., \textit{Nonlocal theories of gravity: the flat space propagator}, arXiv:1302.0532 [gr-qc].
	 
	 \bibitem{Biswas_dS_AdS}
	Biswas, T., Koshelev, A.S. and Mazumdar, A., \textit{Consistent higher derivative gravitational theories with stable de Sitter and anti–de Sitter backgrounds}, \textit{Phys. Rev. D} \textbf{95} (2017) 043533.
	
	\bibitem{Biswas_dS_AdS_2}
	Biswas, T., Koshelev, A.S. and Mazumdar, A., \textit{Gravitational theories with stable (anti-)de Sitter backgrounds}, arXiv:1602.08475 [hep-th].
	    
	\bibitem{Biswas_Eq_Motion}
	Biswas, T., Conroy, A., Koshelev, A.S. and Mazumdar, A., \textit{Generalized ghost-free quadratic curvature gravity}, \textit{Class. Quantum Gravity} \textbf{31} (2014) 159501.
	
	\bibitem{Accioly-Prescrip}
	Accioly, A., Helayel-Netto, J., Barone, F.E. and Herdy, W., \textit{Simple prescription for computing the interparticle potencial energy for D-dimensional gravity systems}, \textit{Class. Quantum Gravity} \textbf{32} (2015) 035021.
    
    \bibitem{Accioly:2000}
	Accioly, A., Ragusa, S., Mukai, H., de Rey Neto, E., \emph{Algorithm for Computing the Propagator for Higher Derivative Gravity Theories}, \emph{International Journal of Theoretical Physics} {\bf 39} (2000) 1599.
    
    \bibitem{Conroy}
	Conroy, A., Mazumdar, A., Talaganis, S., Teimouri, A., \emph{Nonlocal gravity in D dimensions: Propagators, entropy, and a bouncing cosmology}, \emph{Phys.Rev. D} {\bf 92} (2015) 124051.
    
    \bibitem{Quandt}
	Quandt, I., Schmidt, H., \emph{The Newtonian limit of fourth and higher order gravity}, \emph{Astron. Nachr} {\bf 312} (1991) 97.
    
    \bibitem{NMG}
	Bergshoeff, E.A., Hohm, O. and Townsend, P.K., \textit{Massive Gravity in Three Dimensions}, \textit{Phys. Rev. Lett.} \textbf{102} (2009) 201301.
    
    \bibitem{Mazumdar:UV}
	Talaganis, S., Tirthabir, B., Mazumdar, A., \emph{Towards understanding the ultraviolet behavior of quantum loops in infinite-derivative theories of gravity}, \emph{Class. Quantum Gravity} {\bf 32} (2015) 215017.
    
    \bibitem{Ohta}
	Muneyuki, K. and Ohta, N., \textit{Unitarity versus Renormalizability of Higher Derivative Gravity in 3D}, \textit{Phys.Rev. D} \textbf{85} (2012) 101501.
    
    \bibitem{Ohta_unitarity}
	Ohta, N., \textit{A complete classification of higher derivative gravity in 3D and criticality in 4D}, \textit{Class. Quantum Gravity} \textbf{29} (2012) 015002.
	
	\bibitem{Anselmi-1}
	Anselmi, D. and Piva, M., \textit{A new formulation of Lee-Wick quantum field theory}, \textit{JHEP} \textbf{1706} (2017) 066.
	
	\bibitem{Anselmi-2}
	Anselmi, D. and Piva, M., \textit{Perturbative unitarity of Lee-Wick quantum field theory}, \textit{Phys. Rev. D} \textbf{96} (2017) 045009.
	

\end{thebibliography}
\end{document}